\documentclass[manuscript]{aastex}

\shorttitle{DustPedia}
\shortauthors{Davies et al.}

\begin{document}


\title{DustPedia  -  A Definitive Study of Cosmic Dust in the Local Universe}




\author{J. I. Davies\altaffilmark{1}, M. Baes\altaffilmark{2}, S. Bianchi\altaffilmark{3}, A. Jones\altaffilmark{4}, S. Madden\altaffilmark{5}, M. Xilouris\altaffilmark{6}, M. Bocchio\altaffilmark{4}, V. Casasola\altaffilmark{3}, L. Cassara\altaffilmark{6}, C. Clark\altaffilmark{1}, I. De Looze\altaffilmark{2, 10}, R. Evans\altaffilmark{1}, J. Fritz\altaffilmark{9}, M. Galametz\altaffilmark{11}, F. Galliano\altaffilmark{5}, S. Lianou\altaffilmark{5}, A. V. Mosenkov\altaffilmark{2,7,8}, M. Smith\altaffilmark{1}, S. Verstocken\altaffilmark{2}, S. Viaene\altaffilmark{2}, M. Vika\altaffilmark{6}, G. Wagle\altaffilmark{4} and N. Ysard\altaffilmark{4}}
\affil{$^{1}$School of Physics and Astronomy, Cardiff University, Cardiff, CF243YB, UK}
\affil{$^{2}$Sterrenkundig Observatorium, Department of Physics and Astronomy, Universiteit Gent
Krijgslaan 281 S9, B-9000 Gent, Belgium.}
\affil{$^{3}$INAF-Osservatorio Astrofisico di Arcetri, Largo Enrico Fermi 5, I - 50125 Florence Italy.}
\affil{$^{4}$CNRS, Institut d'Astrophysique Spatiale, Paris, France.}
\affil{$^{5}$CEA/DSM/IRFU/Service d'Astrophysique, Astrophysique des Interactions Multi-eschelles (UMR 7158)
CEA, Saclay, Orme des Merisiers batiment 709, 91191 Gif-sur-Yvette, France.}
\affil{$^{6}$National Observatory of Athens, Institute for Astronomy, Astrophysics, Space Applications and Remote Sensing, Ioannou Metaxa and Vasileos Pavlou GR-15236, Athens, Greece.}
\affil{$^{7}$St.Petersburg State University, Universitetskij pr. 28, 198504, St. Petersburg, Stary Peterhof, Russia.}
\affil{$^{8}$Central Astronomical Observatory of RAS, Pulkovskoye chaussee 65/1, 196140, St. Petersburg, Russia.}
\affil{$^{9}$Instituto de Radioastronom\'\i a y Astrof\'\i sica (IRyA-UNAM), Antigua Carrettera a P\'atzcuaro, 8701, Morelia, Michoac\'an, Mexico.}
\affil{$^{10}$Dept. of Physics and Astronomy, University College London, Gower Street, London WC1E 6BT, UK.}
\affil{$^{11}$European Southern Observatory, Karl-Schwarzschild-Str 2,�D-85748 Garching, Germany�}



\begin{abstract}
The European Space Agency has invested heavily in two cornerstones missions; {\it Herschel} and {\it Planck}. The legacy data from these missions provides us with an unprecedented opportunity to study cosmic dust in galaxies so that we can answer fundamental questions about, for example: the origin of the chemical elements, physical processes in the interstellar medium (ISM), its effect on stellar radiation, its relation to star formation and how this relates to the cosmic far infrared background. In this paper we describe the DustPedia project, which is enabling us to develop tools and computer models that will help us relate observed cosmic dust emission to its physical properties (chemical composition, size distribution, temperature), to its origins (evolved stars, super novae, growth in the ISM) and the processes that destroy it (high energy collisions and shock heated gas). To carry out this research we will combine the {\it Herschel/Planck} data with that from other sources of data, providing observations at numerous  wavelengths ($\le 41$) across the spectral energy distribution, thus creating the DustPedia database. To maximise our spatial resolution and sensitivity to cosmic dust we limit our analysis to 4231 local galaxies ($v<3000$ km s$^{-1}$) selected via their near infrared luminosity (stellar mass). To help us interpret the data we have developed a new physical model for dust (THEMIS), a new Bayesian method of fitting and interpreting spectral energy distributions (HerBIE) and a state-of-the-art Monte Carlo photon tracing radiative transfer model (SKIRT). In this the first of the DustPedia papers we describe the project objectives, data sets used and provide an insight into the new scientific methods we plan to implement.
\end{abstract}



\section{Introduction}
In this paper we describe a major collaborative project to enable a much better understanding of cosmic dust and particularly how it influences both physical processes in the interstellar medium and the observations we make. The DustPedia\footnote{DustPedia is a collaborative focused research project supported by  European Union Grant 606847 awarded under the FP7 call. Further information can be found at www.dustpedia.com.} project aims to bring together expertise in various aspects of the study of cosmic dust to develop a coherent interpretation of recent state-of-the-art observations, particularly those that are now available after the successful {\it Herschel Space Telescope} mission (Pilbratt et al. 2010). We have constructed from the {\it Herschel Science Archive} (HSA) a sample of nearby galaxies (within $\sim$3000 km s$^{-1}$ or about 40 Mpc), to study at far-infrared wavelengths. These observations will also be combined with data that is available at many other wavelengths from numerous other databases, thus extending our studies from the ultraviolet to the radio. The far-infrared emission we detect using {\it Herschel} arises mainly from cosmic dust in the interstellar medium that is heated by galactic stars, hence the name of our project - DustPedia.

Cosmic dust forms by nucleation and growth from the vapour phase in the cool atmospheres of low mass stars as they come to the end of their lives and probably also in the gas ejected from supernovae as more massive stars expire. Once deposited into the interstellar medium the dust grains are subject to various physical processes that allow them to grow via the accretion of atoms and molecules and disintegrate in shock heated gas or via high-energy photon or cosmic ray processing. The dust is composed of a mixture of carbonaceous and amorphous silicate grains with a size distribution governed by the growth and destruction mechanisms (sizes of order 0.01-1.0$\mu$m). A comprehensive discussion of dust in the inter-stellar medium can be found in Whittet (2002) and Tielens (2005).

A large part of our project is to study the dust formation process, dust evolution and destruction, to eventually construct a model able to explain the diversity of observations we make. Our observations sample a wide range of physical conditions (density, temperature and composition) within the inter-stellar medium of galaxies and as such our models need to be versatile.

There are three primary physical mechanisms that enable us to detect cosmic dust in galaxies. Firstly, dust absorbs and scatters radiation from the stars, which not only causes the extinction of the light, but also causes the stellar spectrum to become "redder". Secondly, the alignment of dust grains in the Galactic magnetic field leads to the polarisation of starlight. Finally,
because the absorbed stellar light heats the dust it subsequently radiates some of its energy away. Dust temperatures range from about 10-100K and so the energy radiated from dust is emitted predominantly in the mid-infrared and sub-mm part of the electromagnetic spectrum (wavelengths of about 10$\mu$m to 1mm). Imprinted on this spectrum are signatures of dust composition, structure and chemistry, which provide input into a realistic physical dust model. In addition the combination of the observed absorption and scattering with that of the measured emission, via a radiative transfer code, provides other important measurements of the physical properties of the dust grains (for example, dust emissivity). A comprehensive review of how dust observations can be related to the properties of inter-stellar dust grains can be found in Draine (2003).

Dust extinction and polarisation are effects imprinted by dust on the radiation from stars, while the only direct measure of the dust itself comes from the radiation it emits. The detection and measurement of this radiation had to await the arrival of space telescopes, because the majority of the cosmic far-infrared and sub-mm radiation is efficiently absorbed by molecules in the Earth's atmosphere. The first far infrared space telescope ({\it IRAS}, Neugebauer et al.1984) revolutionised our ideas about the physical properties of cosmic dust, just how much dust there was and how important the dust is in governing physical processes in the interstellar medium (Rice et al. 1988 and references therein). For a typical galaxy like the Milky Way from one third to a half of the radiation produced by its stars is subsequently re-processed through cosmic dust (Popescue and Tuffs 2002, Davies et al. 2012, Davies et al. 2013). We now know of other, more extreme galaxies, in which 99\% of the stellar radiation is reprocessed in this way (Sanders and Mirabel 1996). Subsequent pre-{\it Herschel} space missions ({\it ISO}, Kessler et al. 1996, {\it Spitzer}, Werner et al., 2004) have greatly extended our understanding of cosmic dust in galaxies providing detailed information on the physical properties of the dust and how it is spatially distributed (see for example Alton et al. 1998, Munoz-Mateos et al. 2011, Galametz et al. 2011). In addition, a wealth of information has been gleaned about dust in our own galaxy by telescopes primarily designed to observe the cosmic microwave background i.e. {\it COBE}, {\it WMAP}, and {\it PLANCK} (Sodroski et al. 1997, Bennett et al. 2003, Green et al. 2015). 

The {\it Herschel Space Observatory} (Pilbratt et al. 2010) has hugely advanced our understanding of cosmic dust, because with a mirror diameter of 3.5m its collecting area is 5-6 times larger than any of the previous far-infrared telescopes - giving both improved sensitivity and spatial resolution. One of the major discoveries from previous far-infrared missions was that the cosmic dust was somewhat colder than expected (Alton et al. 1998, Galametz et al. 2011, Galliano et al. 2003, 2005), so the {\it Herschel} instruments were also designed to look at a previously un-explored part of the electromagnetic spectrum between the far-infrared and sub-mm (250-500$\mu$m). This was in addition to the spectral regions previously explored with smaller telescopes (70-160$\mu$m). A major part of the DustPedia project is to exploit the unique capabilities of the data in the {\it Herschel} legacy archive and use them, not just to constrain the physical dust model, but to explore broader science issues as well (see section 6).

Recent mid-infrared to sub-mm observations of cosmic dust ({\it WISE}, Wright et al. 2010, {\it Spitzer}, Werner et al., 2004, {\it Herschel}, Pilbratt et al. 2010) and their interpretation can be considered important for four primary reasons. 
\begin{enumerate}
\item As a depository of metals the dust content of a galaxy at face value is a measure of how far along the evolutionary path a galaxy has progressed (Edmunds and Eales 1998, Dunne et al. 2003, Davies et al. 2014). \item Cosmic dust plays an important role in many of the physical processes that regulate the evolution of galaxies. For example, it provides opacity so that giant clouds of gas collapsing under gravity can heat up to temperatures sufficient for stars to form and nucleosynthesis to start. It is on the surface of dust grains that molecular hydrogen, which is the crucial gaseous ingredient for star formation, forms. 
\item Dust traces other physical processes and galaxy constituents that are not so easily measured. For example far infrared emission from galaxies is closely related to the rate at which stars form (Bell 2003, Calzetti et al. 2007, Davies et al. 2014, ) and the relatively easily measured mass of dust relates closely to the difficult to measure mass of molecular hydrogen (Galametz et al. 2011, Magrini et al. 2011, Eales et al. 2012, Remy-Ruyer et al. 2013). 
\item Dust can greatly affect what you measure at other wavelengths. The ultra-violet emission of hot young stars for example is greatly attenuated by dust and may lay hidden and the reddening effect can mislead us in our determination of the ages of stellar populations. 
\end{enumerate}

On cosmological scales the diffuse far infrared background can be used to infer the far infrared luminosity and dust temperature of distant galaxies and as a measure of the star formation history of the Universe (Puget et al. 1996). However, the interpretation of this background depends very much on an accurate local measurement of the far infrared luminosity density and temperature. We will use the DustPedia sample to measure the local far infrared luminosity density in all available bands and then use this, via a suitable calibration (Davies et al. 2014) to measure the local star formation rate density. In addition, dust extinction through the Universe may noticeably influence our observations of the most distant objects (Menard et al. 2010). We will be able to provide the most up-to-date analysis of the dust extent and column density profile of galaxies and so use this to revisit the question of the line of sight extinction and reddening to distant objects. 

\subsection{The Importance of Studying Nearby Galaxies}
Many observations of galaxies over large look back times correspond very well with cosmological models of galaxy and larger scale structure formation. This has led to the wide spread belief that the current cosmological model is broadly correct. Although there is good agreement over large spatial scales there are some challenging disagreements between theory and observation when one looks over smaller scales and particularly at the properties of nearby galaxies. The distributions of galaxy mass and size, their locations within larger scale structures in the Universe and their star formation histories as a function of galactic mass are all examples of disparity with the currently favoured model. In particular Peebles and Nusser (2010) stress the importance of nearby galaxies if we want to understand the detailed processes of galaxy evolution and hence develop a complete model of how galaxies change with time. They specifically say that  ".......nearby galaxies offer rich and still far from completely explored clues to a better picture of how galaxies form.". The reason of course is that nearby galaxies can be studied in far greater detail than those that lie at the edge of the Cosmos, and importantly typical cosmological surveys have covered such small areas of sky that they do not sample the local population very well if at all. Observations of cosmic dust address many aspects of the current galaxy evolutionary model i.e. star formation rate, growth of the metal abundance, loss of metals in galactic winds, physical processes in the interstellar medium etc. and so offer the potential for a much better understanding. Within the DustPedia project we will carry out five specific tasks, utilising nearby galaxies, that will provide important inputs into current evolutionary models:
\begin{enumerate}
\item Measure the complete UV-mm/radio spectral energy distribution (SEDs) for a large number ($>$1000) of galaxies, and for different environments within individual galaxies. 
\item Interpret the galaxy SEDs using radiative transfer and full SED models, to derive stellar, gas and dust properties, star formation rates and histories as a function of morphological type. 
\item Determine how the dust NIR-mm/radio SED evolves throughout the Universe and how this is related to the underlying dust properties. 
\item Develop a dust evolution model that is consistent with the SEDs of galaxies of different morphological types and determine the primary sources and sinks for cosmic dust. 
\item Derive dust mass functions to the lowest-possible luminosities and masses and compare these with cosmological surveys and the cosmic far infrared background.
\end{enumerate}

\section{The Dustpedia data set}
\begin{figure}
\centering
\includegraphics[width=14cm]{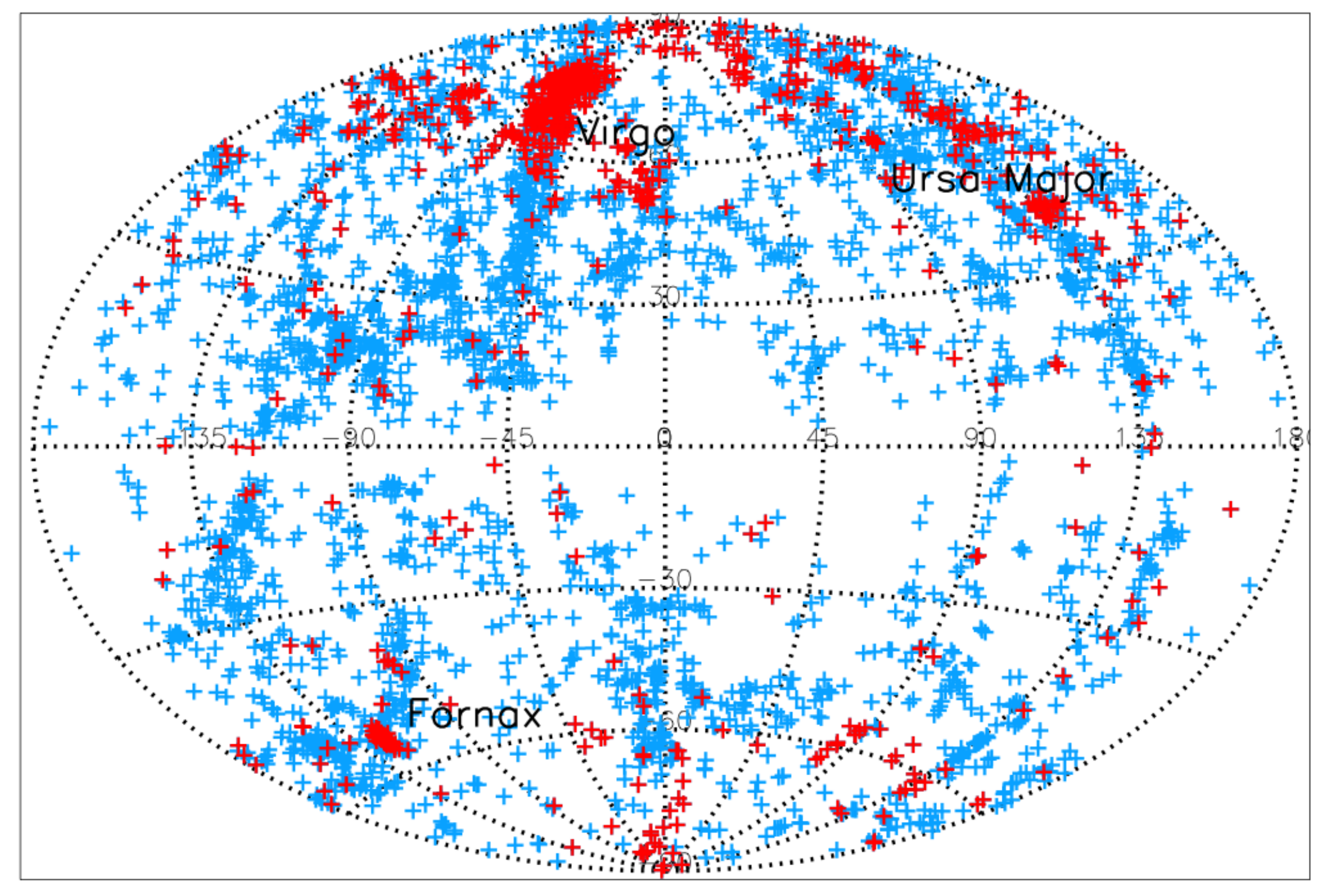}
\caption{The spatial distribution of galaxies in Galactic coordinates but plotted so that for $l>180^{0}$ the plotted value is $l-360^{0}$. Blue - galaxies selected from {\it WISE}. Red - {\it WISE} galaxies with a {\it Herschel} observation. The Virgo and Fornax clusters are labeled along with Ursa Major where there is also a concentration of nearby galaxies.}
\label{fig:gal_dist}
\end{figure}

Our intention was to select a sample of nearby galaxies, that included a subset that were resolved by {\it Herschel} ($D_{25}>5$ arc min), but also contained objects that were far enough away that we sampled the larger scale structure and hence different galactic environments. We also wanted to primarily select objects according to their stellar mass, the best proxy for which is near infrared flux density (Eskew et al. 2012, Meidt et al. 2014). To satisfy these criteria the initial definition of the DustPedia sample was that it should consist of all galaxies observed by {\it Herschel} that lie at recessional velocities of $<3000$ km s$^{-1}$, with optical diameters $>1$ arc min and a {\it WISE} 3.4$\mu$m signal-to-noise ratio (SNR) $>5$. The velocity restriction means that we include galaxies that are 'local' yet still reside in different environments. For example this selection includes the Virgo and Fornax clusters and galaxies in the super-galactic plane that consists of the Virgo southern extension and connecting galaxies from Virgo to the less rich Ursa Major cluster (Fig.~\ref{fig:gal_dist}). The size constraint means that all of our sources are extended in every {\it Herschel} band, even if they're not fully resolved. The {\it WISE} flux provides for our required stellar mass selection.

\begin{table}[h]
\begin{center}
\begin{tabular}{lc} \hline
Selection Criterion & Number \\ \hline
LEDA & 4259 \\
LEDA+WISE$_{5\sigma}$ & 4231 \\
LEDA+WISE$_{5\sigma}$+PACS  & 829 \\
LEDA+WISE$_{5\sigma}$+SPIRE  & 907 \\
LEDA+WISE$_{5\sigma}$+PACS+SPIRE  & 798 \\
LEDA+WISE$_{5\sigma}$+PACS/SPIRE+Inspection  & 876 \\
\end{tabular}
\caption{Summary of the numbers of identified galaxies when applying the various DustPedia sample selection criteria.}
\end{center}
\end{table}

We initially interrogated the LEDA\footnote{http://leda.univ-lyon1.fr/} database (Makarov et al. 2014) from which we selected 4259 galaxies that had $v_{rad}<3000$ km s$^{-1}$ and $D_{25}>1$ arc min, where $v_{rad}$ is the measured radial velocity either from an optical or 21cm spectrum and $D_{25}$ is the major axis diameter measured at the 25th blue magnitude per sq arc sec isophote. The {\it WISE} All-Sky Release Explanatory Supplement\footnote{
$http://wise2.ipac.caltech.edu/docs/release/allsky/expsup/sec6_3a.html$}
states that at 3.4$\mu$m {\it WISE} has a 95\% sky coverage at a 5$\sigma$ sensitivity limit of 0.039 mJy (19.91 mag$_{AB}$, 17.23 mag$_{Vega}$). When matched to the LEDA data the {\it WISE} selection criterion only very slightly reduced the sample size from 4259 to 4231.

It was also decided that studying the
very large angular size Local Group galaxies M31 (and its companions), M33, LMC and SMC falls outside the scope of the project,
given the very different nature of working with such extended systems. We also excluded any galaxies along the line of sight to these large angular size galaxies because this makes them particularly difficult to analyse.

\section{The {\it Herschel} data and its reduction}
\subsection{Observation identification}
We have interrogated the {\it Herschel Science Archive} (HSA) to look for photometer (not spectra) observations of our LEDA and {\it WISE} selected targets. The two relevant {\it Herschel} instruments are The Photoconductor Array Camera and Spectrometer (PACS) (observations at 70, 100, 160$\mu$m, Poglitsch et al., 2010) and the Spectral and Photometric Imaging Receiver (SPIRE) (observations at 250, 350, 500$\mu$m, Griffin et al., 2010). The observations are obtained from all of the numerous science projects undertaken by the Herschel observatory. Of the 4,231 DustPedia candidate objects, 907 have been observed by SPIRE, whilst 829 have been observed by PACS. The vast majority, 798, have been observed by both instruments; 109 SPIRE data only, and 31 PACS data only. According to the HSA a total of 938 have been observed by either PACS and/or SPIRE, but upon visual inspection only 876 of these had useful data\footnote{Galaxies were excluded if an appreciable portion of their area fell outside the area of the map or if they were in the turn around region of a scan map.}, this constitutes our primary DustPedia sample. Table 1 summarises the numbers of sources found when applying the various sample
selection criteria. 
\begin{figure}
\centering
\includegraphics[width=8cm]{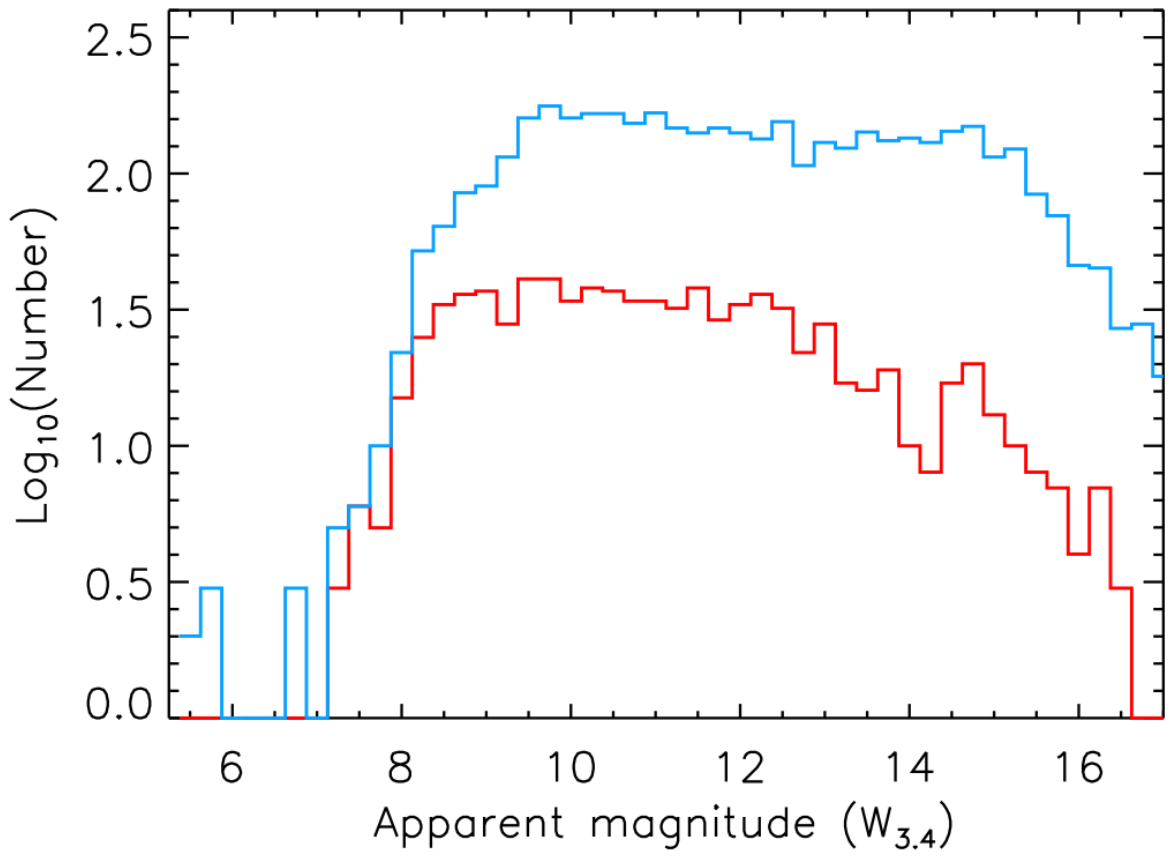}
\includegraphics[width=8cm]{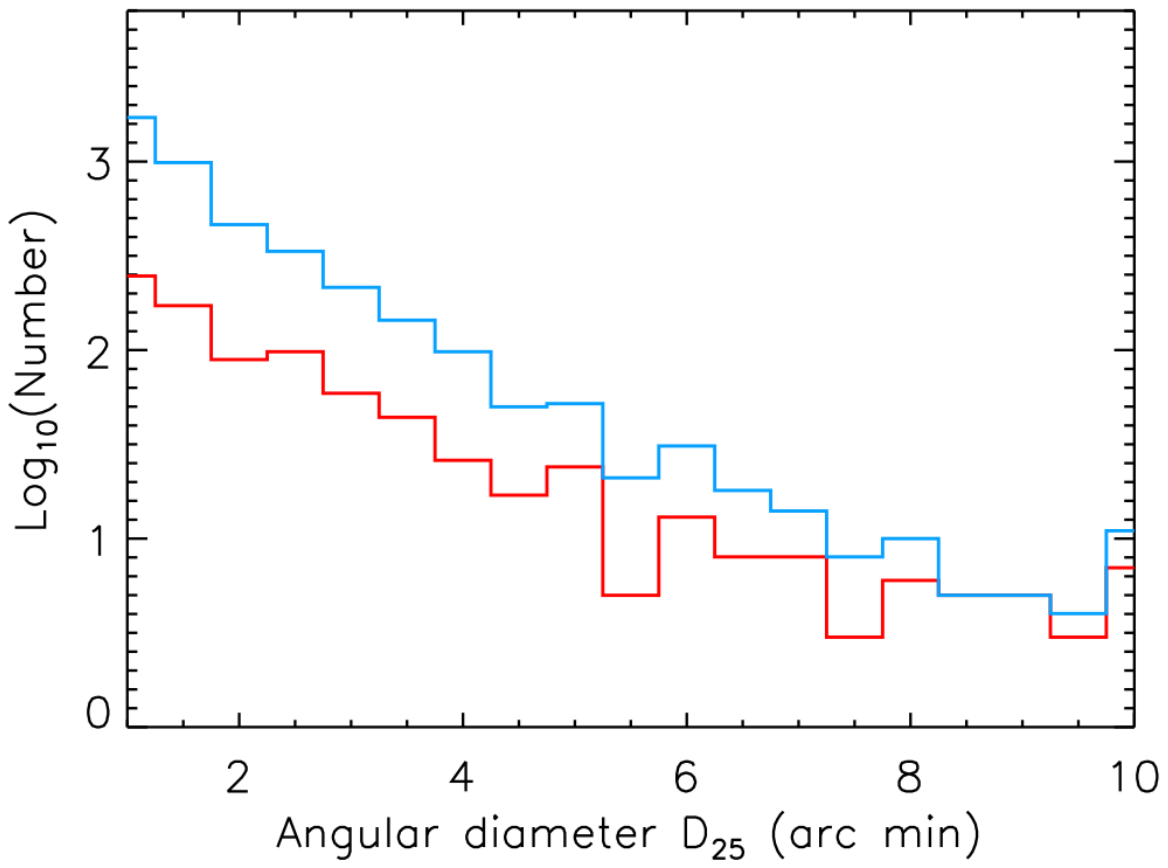}\\
\vspace{-4.0cm}
\includegraphics[width=8cm]{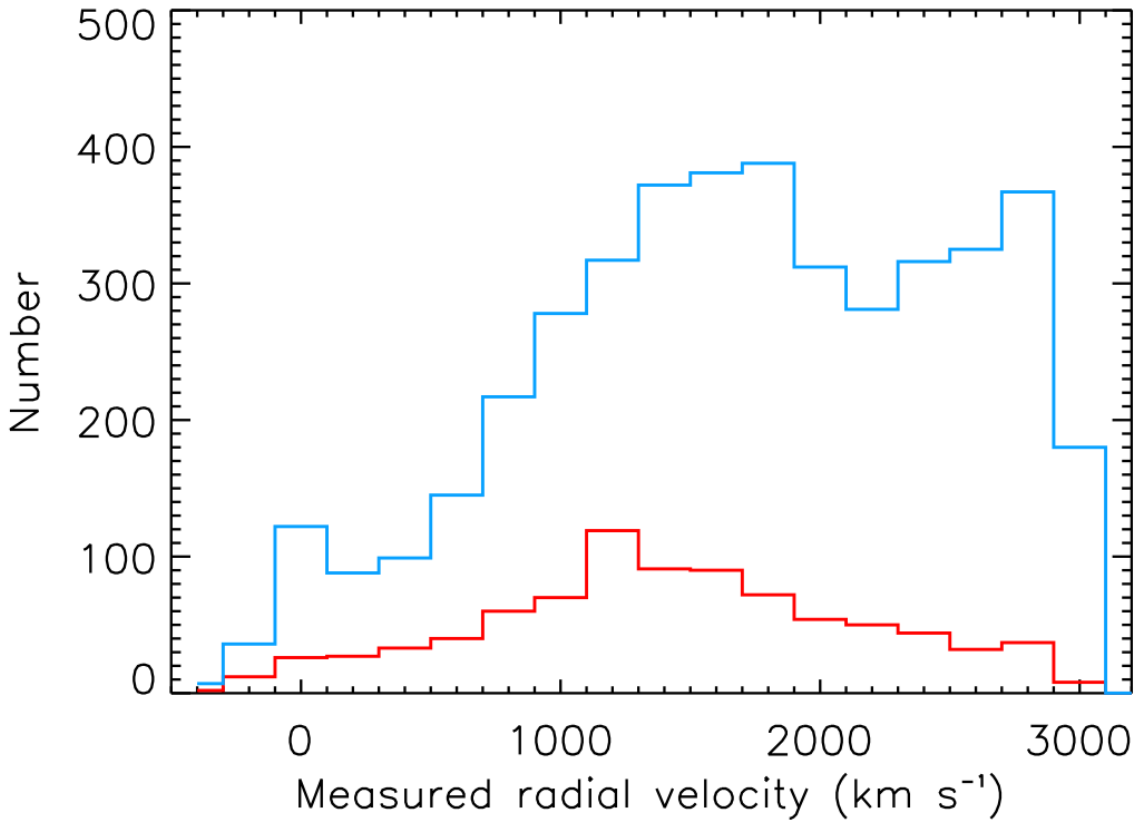}
\includegraphics[width=8cm]{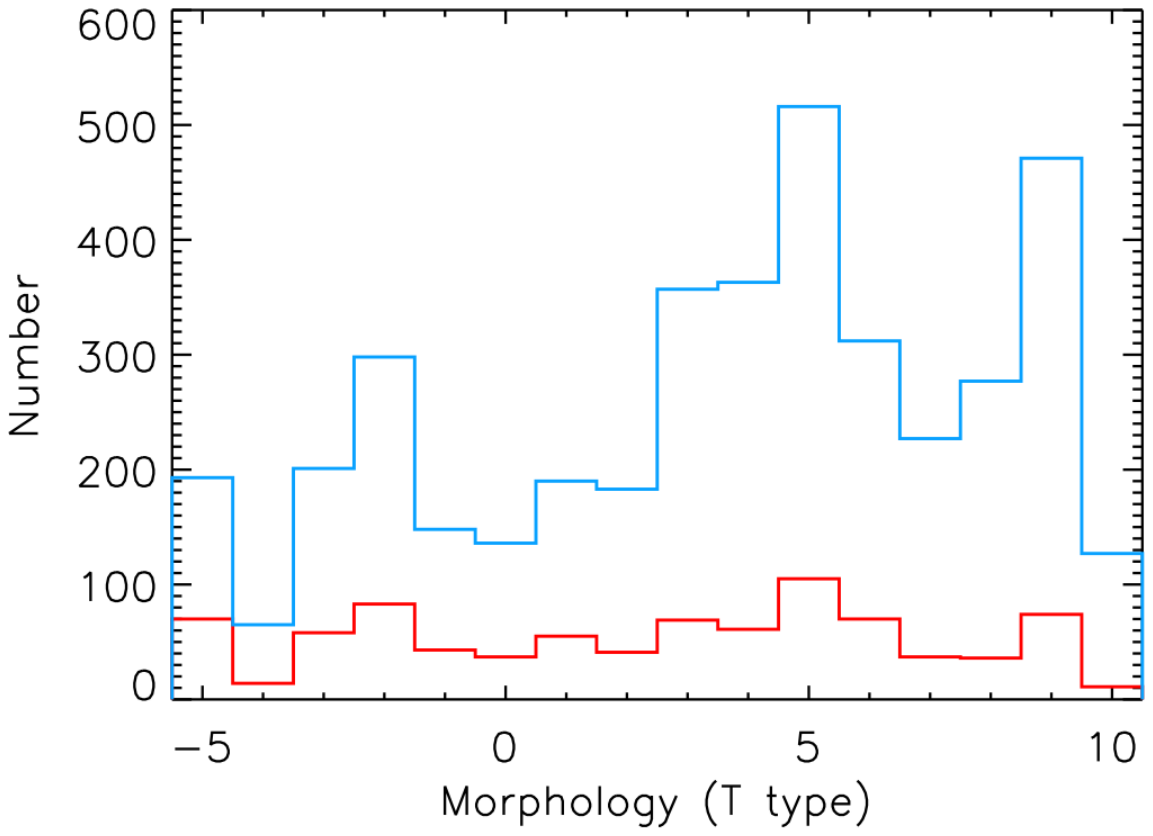}\\
\vspace{-2.0cm}
\caption{The above figures compare the properties of the galaxies selected from {\it WISE} (blue) with those that also have a {\it Herschel} detection (red). Top left - {\it WISE} 3.4$\mu$m apparent magnitude. Top right - Optical angular diameter. Bottom left - measured radial velocity. Bottom right - morphological type.}
\label{fig:sample_prop}
\end{figure}

In Fig.~\ref{fig:sample_prop} we compare the properties of the galaxies selected using LEDA and {\it WISE} (blue) with those that additionally have a detection by {\it Herschel} (red). In the main the {\it Herschel} detected galaxies appear to broadly follow the distributions of the primary (LEDA and {\it WISE}) selected galaxies. However, using a Kolmogorov-Smirnov two sided test we can reject the hypothesis that the {\it WISE}(LEDA) and {\it Herschel} data are drawn from the same parent distributions for the parameters shown in Fig.~\ref{fig:sample_prop} . This entirely foreseeable problem of course arises, because of the rather arbitrary way in which the various un-coordinated Herschel projects we have obtained data from, have sampled the local galactic population. To overcome this problem our intention is to use the data we have on the  detected galaxies to construct advanced predictors (principle component analysis, machine learning) to infer properties for non-detected galaxies. In this way we hope to apply much of our analysis to the entire 4231 {\it WISE}(LEDA) galaxy sample. 

The morphological mix of the {\it Herschel} sample is also shown in Fig.~\ref{fig:sample_prop} and because we will consider galaxies of different morphology separately (section 6) we also give the numbers of each type in Table 2 - as expected, because of their generally larger dust mass, the numbers are dominated by late type galaxies. About 25\% of the {\it Herschel} sample galaxies lie in the Virgo cluster and about 5\% in Fornax.

\begin{table}[h]
\begin{center}
\begin{tabular}{cccc} \hline
Type & Number \\ \hline
Early ($T \le -4$) & 72 \\
S0 ($-4 < T \le 0$) & 203 \\
Spiral ($0 < T \le 7$) & 433 \\
Irr/dwarf ($7 < T$) & 159 \\
\end{tabular}
\caption{Numbers of galaxies of different morphological types in the {\it Herschel} sample.}
\end{center}
\end{table}

\subsection{SPIRE data reduction}
The SPIRE data reduction was carried out in several stages. In the first stage, the raw Level-0 data (acquired from the HSA) was processed up to Level-1 (photometrically and astrometrically calibrated timelines) using HIPE v13\footnote{HIPEv13 was the then current release of the {\it Herschel} Interactive Processing Environment (Ott, 2010): http://www.cosmos.esa.int/web/herschel/hipe-download.}. The Level-1 data was then processed to Level-2 using the BriGAdE (Bright Galaxy Adaptive Element, Smith, 2012a) software; BriGAdE operates in the same manner as the standard HIPEv13 Level-1 to Level-2 pipeline, but with two key differences.

The first key feature of BriGAdE is that it identifies and removes "jumps" in the thermistor timeline\footnote{The cause of these "jumps" is not clear, but it is something in the instrument electronics which causes an instantaneous DC offset in the detectors.}. This is in contrast to the standard HIPEv13 processing, which simply switches to alternate thermistor timelines, which may themselves include jumps, which then often leads to conspicuous image artefacts. We manually inspected our timelines for cosmic ray glitches, bolometer jumps, and thermistor jumps, in addition to the automated identification process provided by HIPE v13.
\begin{figure}
\centering
\vspace{-11.0cm}
\includegraphics[width=18cm]{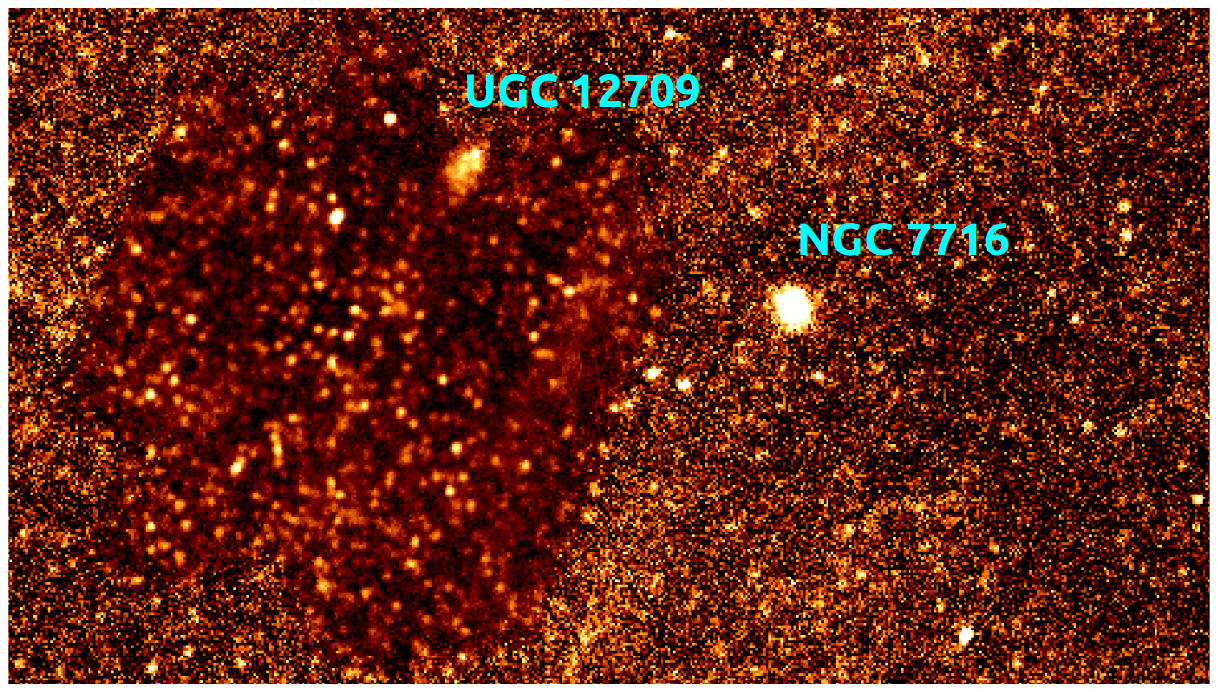}
\vspace{-9.0cm}
\caption{SPIRE 250 $\mu$m map containing DustPedia galaxies UGC 12709 and NGC 7716, illustrating how all available SPIRE data for DustPedia galaxies was combined. The whole image is covered by shallow HerMES ({\it Herschel} Multi- tiered Extragalactic Survey, Oliver et al., 2012) data, whilst the left of the image also has coverage from super-deep HLS (Herschel Lensing Survey, Egami et al., 2010) data, producing a small region with ultra-deep coverage that serendipitously covers UGC 12709.}
\label{fig:spire_ugc12709}
\end{figure}
The second key feature of BriGAdE is that the individual bolometer and thermistor scan leg timelines are combined to form contiguous timelines. By using contiguous timelines for entire observations, it is possible to constrain and remove the instrumental thermal drift in a way that is superior to the standard HIPEv13 processing. The contiguous array thermistor timelines (with jumps and other artefacts already removed) are directly compared to the contiguous bolometer time lines, to establish which thermistor provides a superior fit to the instrumental thermal drift.  In order to constrain the instrumental thermal drift, BriGAdE has to identify and mask bright sources from the timeline; in observations containing large amounts of galactic cirrus, this is not possible. In observations of such regions, our reduction therefore proceeds in a manner essentially identical to the normal HIPE v13 processing.

Observations that covered the same patches of sky were identified and combined (see for example Fig.~\ref{fig:spire_ugc12709} ), with an exception for observations of the Virgo Cluster. The sheer quantity of data in the direction of Virgo meant that combining all available observations was computationally prohibitive. As such, in regions covered by HeViCS (the {\it Herschel} Virgo Cluster Survey, Davies et al. 2010), we made use of the HeViCS data alone. This should have no significant impact on the quality of our final data products, as the HeViCS data is very deep for a large-area nearby galaxy survey, with noise of 6.6 mJy beam$^{-1}$ at 250 $\mu$m (Auld et al., 2013); sufficiently deep that it is actually dominated by confusion noise, which is of order 5.8 mJy beam$^{-1}$ at 250 $\mu$m (Nguyen et al., 2010).
The finalised timelines were run through the HIPE v13 de-striper. The improved thermal drift correction provided by BriGAdE, and the fact that we combine co-local observations, allows the de-striper to operate more effectively than it otherwise would.

The final maps were produced from the reduced data using the HIPEv13
naive map-maker, with pixel units of Jy pix$^{-1}$, and pixel sizes of 6, 8, and 12 arc sec
at 250, 350, and 500 $\mu$m respectively, each corresponding to approximately one-third of the beam width at each wavelength. The maps do not include the {\it Planck} flux level offset found in the standard HIPEv13 reductions. For DustPedia sources found in maps larger than one degree in size, we produced $1^{o} \times 1^{o}$ cutouts centred on the source, whilst for DustPedia sources found in maps smaller than $1^{o}$ the entire map represents the final data product. Having such large maps is beneficial for SPIRE observations in particular, where galactic cirrus can be a significant contaminant. Having a large region of sky over which to characterise the cirrus makes it possible to subtract it (Auld et al., 2013, Bianchi et al., 2016), and allows an accurate determination of the cirrus contribution to the aperture noise, and hence total flux uncertainty, of the photometry of a given source.

Visual inspection of the reduced data revealed that 63 DustPedia candidate objects did not have usable SPIRE coverage (being located in scan turnaround regions). So, in total, the final SPIRE reductions produced coverage for 844 of the DustPedia candidate objects.

\subsection{PACS data reduction}
The PACS data reduction was performed in two stages. The first step (in which the raw Level-0 data is processed up to Level-1) includes the creation of calibrated timelines, coordinate assignment, masking of bad pixels, removal of calibration blocks, and flat-field correction; this was performed with HIPE v13. The second step includes deglitching, correction for 1/f noise, and map making; this was done using the standalone software Scanamorphousv24 (Roussel, 2013).
The Scanamorphous processing took account of whether the observations
in question had been taken in scan map mode, mini map mode, or parallel
mode. The final DustPedia PACS maps are in units of Jy pix$^{-1}$ with pixel
sizes of 2, 3, and 4 arc sec at 70, 100, and 160 $\mu$m respectively. In
keeping with our SPIRE maps, these pixel sizes correspond to approximately one-third of the beam width. For DustPedia galaxies found in very large fields (such as HeViCS or H-ATLAS), cutouts were produced of $0.35^{o} \times 0.35^{o}$ in size, centred on the target.
All objects observed with PACS have coverage at 160 $\mu$m. However, the design of the instrument means that coverage in the 70 and 100 $\mu$m bands is mutually exclusive for a given observation. The majority of DustPedia galaxies have 160 and 100 $\mu$m coverage, whilst a number (where PACS observations were repeated) have coverage in all three bands.
Unlike with SPIRE, it is not possible to combine PACS observations made in different observing modes. This is primarily because the size, elongation, and overall morphology of the PACS point spread function (PSF) varies considerably between observing modes. As such, for DustPedia sources that have been observed by PACS in more than one observing mode or program, separate maps were produced using the data from each. 
The superior set of observations was decided for each source on a case-by-case basis, preferring maps with deeper coverage, smaller scan 
speed, and larger sky extent. In most of these cases, parallel mode 
observations - generally being shallower than dedicated observations - 
were used only to retrieve the 100um maps for objects with 70/160 
scan maps only. In total the PACS data reductions produced coverage for 829 of the DustPedia candidate objects.

\section{Ancillary data}
For the 876 DustPedia galaxies with {\it Herschel} imaging coverage, ancillary imaging data has been gathered from 7 telescopes that have observed large numbers of nearby galaxies. This provides combined coverage of an additional 32 bands; although of course not all galaxies have been observed in all bands. Multi-wavelength data is available for the majority of the primary (observed by {\it Herschel}) sample as follows; GALEX (GR6/7, Bianchi et al., 2014) coverage is available in at least one band for 824 (95\%), SDSS (DR9, Ahn et al., 2012) coverage is available for 652 (75\%), and {\it Spitzer} coverage is available in at least one band for 821 (95\%), along with the all-sky coverage provided by {\it DSS} in the optical, {\it 2MASS} (Skrutskie et al., 2006) in the NIR, and {\it WISE} (Wright et al., 2010) in the MIR. At 22$\mu$m {\it WISE} has a 95\% coverage at a 5$\sigma$ sensitivity level of 5.1 mJy (8 magAB, 14.6 magVega). Correlating the positions of the DustPedia sources with the Second {\it Planck} Catalogue of Compact Sources ({\it Planck} Collaboration et al., 2015), using a matching radius equal to the {\it Planck} beam FWHM in each {\it Planck} band, returns matches for 439 (51\%) of the sample. However, the shorter {\it Planck} wavelengths have more detections; there are 393 matches with the {\it Planck} 350$\mu$m catalogue, whilst there are only 196 matches with the {\it Planck} 850$\mu$m  catalogue.
When including the {\it Herschel} PACS and SPIRE data, combined with {\it Planck} catalogue fluxes and {\it IRAS} SCANPI measurements, the DustPedia database is able to provide photometry in up to 41 bands\footnote{The availability of data is of course continually changing and we hope to regularly update our data base. AKARI data is an obvious omission that we hope to rectify soon, particularly given the data described in Solarz et al. (2016). Other surveys such as those being carried out using the VST and VISTA telescopes (KIDS, de Jong et al., 2013, VIKING, Edge et al., 2013) will improve optical and NIR coverage over the southern hemisphere. We are currently looking into including data on the gas content and spectral information.}.

For observations of a given target galaxy, in a given band, by a given telescope, all available imagery covering the region of interest was identified. All data was retrieved from the official archive of the telescope in question. These images were then mosaiced to produce a cutout image.
In general, each cutout is a $0.5^{o} \times 0.5^{o}$ box (oriented East-North), centred on the target galaxy. However, for galaxies so extended that they would fill more than 20\% of the width of such a box (i.e. galaxies with angular sizes $>6$ arc min), then the cutout was expanded to $1^{0} \times 1^{0}$; less than 10\% of the DustPedia galaxies with {\it Herschel} coverage are this extended. These larger cutouts provide sufficient sky to allow for aperture noise estimation, background fitting, etc, for even the most extended targets.

There were instances where there were archive observations in a given band within the region of interest, but the location of the target galaxy itself was not covered. To overcome this problem the central $10\times10$ pixels of each  cutout (i.e. the location of the target galaxy) was inspected; if all of them were empty, then the cutout was rejected.
The tool employed to perform the mosaicing was Montage (Berriman et al. 2016). Montage was used to re-grid the images to a common projection, match their background levels, and co-add them; in cases where the images being co-added were of different depths, the contribution of each image to the co-addition was appropriately weighted (using error maps, exposure time information, etc).
Note that the background-matching carried out by Montage is not a background subtraction. Rather, Montage adjusts the level of each image, so that it matches, as closely as possible, those images it overlaps with. 

The image FITS headers contain the standard World Coordinate System (WCS) fields, along with fields that provide a range of additional information: target galaxy, telescope, filter effective wavelength, filter name (if applicable), instrument (if applicable), map units, and the name of the database the original data was acquired from. If there is an error map available, then the header additionally indicates whether the file contains the image map or the error map.
As with the {\it Herschel} maps, the ancillary data cutouts have all been standardised to be in units of Janskys per pixel. 

The archive data in each band is currently at the pixel scale appropriate for that band, with no smoothing carried out, although all the maps have been reprojected to a North-East projection. Pixel-by-pixel comparisons will require the pixel grids and PSFs to be matched, however different projects will have different requirements in this regard. For example some users might want to keep the 500$\mu$m data in order to maximise wavelength coverage, whilst others may leave out the 500$\mu$m and instead only go out to 350$\mu$m, in order to benefit from its improved resolution. There are also DustPedia galaxies for which there is only PACS data, no SPIRE, in which case smoothing to the SPIRE would not be necessary.  Currently we believe that the legacy value of the dataset to the community is better if it is left in a relatively 'raw' form, which will allow users to process it in the way they see fit.

The above data were used to produce a standardised set of cutouts. In total, the ancillary imaging data comprises of 25,120 maps. Of the 876 DustPedia galaxies, 119 (14\%) have the full set of ancillary cutouts, with coverage in all 32 bands. 

\begin{figure}
\centering
\vspace{-8.0cm}
\includegraphics[width=18.0cm]{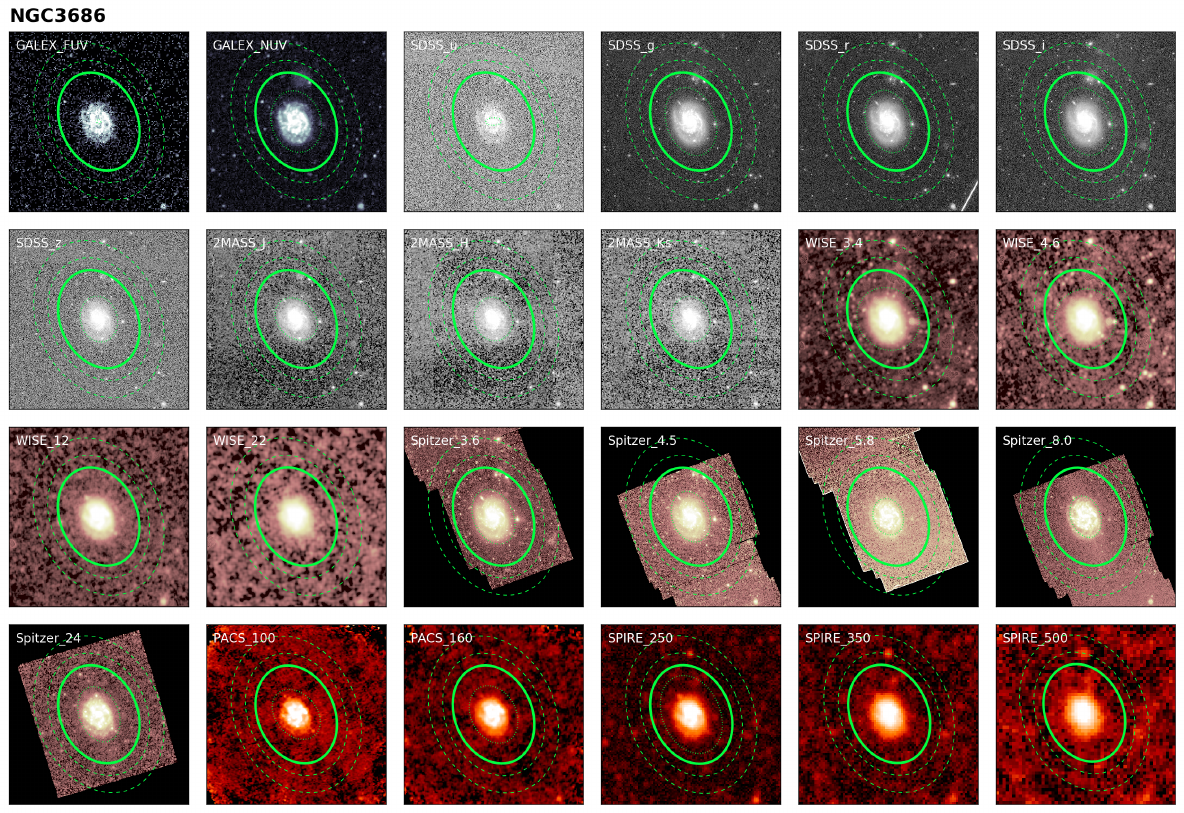}
\vspace{-8.0cm}
\caption{Multi-band images of NGC3686 illustrating the quality of the data available and the apertures used for photometry. The solid green line is the aperture while the dashed lines define the region used to define the background.}
\label{fig:mulitband_phot}
\end{figure}

The entire data have been incorporated into the DustPedia database, which is currently open to consortium members only, but will eventually be accessible by the public at http://dustpedia.astro.noa.gr/. Examples of the available data are shown in Fig. \ref{fig:mulitband_phot}. Users can search the database by galaxy name, Hubble type, velocity, inclination angle and diameter ($D_{25}$). The photometry (see below) in the various bands is currently being incorporated into the database and there will be options to output calibrated spectral energy distributions (SEDs). 

\section{Photometry}
    Using our database imagery, we have performed aperture-matched multi-wavelength photometry for all sources using the CAAPR\footnote{\url{http://github.com/stargrazer82301/CAAPR}} pipeline. The CAAPR pipeline is a development of the photometry pipeline used in Clark et al. (2015) and De Vis et al. (in prep.), and is described in full in Clark et al. (in prep.), where the full photometric data will be presented.

    Initially, foreground stars were removed from the ultra violet to mid-infrared images, using the star-removal functionality of Python Toolkit for SKIRT (PTS\footnote{\url{https://github.com/SKIRT/PTS}}, Camps \& Baes 2015). Then for a given source in a given band, an elliptical aperture is fit to the source, out to the isophote with a signal-to-noise of two; this aperture-fitting is repeated in each available band. The apertures derived for each band were then compared, to find the smallest ellipse that contained every band's aperture. The major and minor axes of this combined aperture were then increased by a factor of 1.2 to define the photometric aperture used in each band. When the final photometry was performed in a given band, this aperture's dimensions are convolved with that band's beam width, to ensure that the photometry in each band is directly comparable. The background-subtraction was performed using a sky annulus centrally positioned on the source.

    To estimate uncertainties on our photometry, we used randomly-located sky apertures placed around each target source. The variation in flux recorded within these apertures incorporates the contributions from instrumental noise, confusion noise, and aperture noise, allowing for measured uncertainties in each band to be directly comparable. Where possible, we used random sky apertures with the same area as the source aperture; however where our maps were of insufficient size to permit this, we used an alternate approach. For these sources we used apertures with a range of smaller sizes to establish the relationship between uncertainty and sky aperture area, and then extrapolated to find what the photometric uncertainty would have been, had it been possible to employ full-sized random apertures.

\section{Dustpedia - science}

In the following sections we describe some of the specific science projects we are undertaking as part of the DustPedia project.

\subsection{SED models to explain the near infrared to sub-mm observations and 
              constrain the physical properties of the dust}

There are several approaches to SED modelling.
Radiative transfer (RT) models (Sect.~6.2) are the most rigorous, as they connect the microscopic grain properties to their macroscopic distribution.
However, a more empirical SED modelling approach has proven to be successful at deriving reliable trends between certain dust properties.
Contrary to RT models, this approach does not contain explicit assumptions about the geometry of the source we are modelling, but simply assumes a general distribution of radiation fields, though mostly of a power-law form:
\begin{equation}
  \frac{dM}{dU} \propto U^{-\alpha} \;\;\;\mbox{for } U_-\le U\le U_+,
\end{equation}
where $U$ is the radiation field intensity and $M$ the dust mass;
$\alpha$, $U_-$ and $U_+$ are usually free parameters.
Similar methods have been implemented by, among others by Dale et al. (2001), Draine and Li (2007) and Galliano et al. (2011).
The significant advantage of this method is that we are able to derive values of the physical parameters independent of the geometry (IR power, dust mass, moments of the radiation field distribution, PAH mass fraction, submm excess, {\it etc.}) with a better precision than RT models. The only limitation in terms of physical diagnostics is that we are consequently not able to extract information about the geometry of the source. 

This approach has proven to be successful in many areas. 
For instance, Remy-Ruyer et al. (2015) performed the most complete study of evolution of the dust-to-gas mass ratio in galaxies, using this model. 
Galliano et al. (2011) were also able to demonstrate, studying the Large Magellanic Cloud, that the grain composition constrained by {\it Herschel}, was more emissive by a factor of $\simeq2-3$ than what was previously assumed. 
This latter discovery has since then been confirmed by {\it Planck} ({\it Planck} collaboration XXIX, 2014) in the Milky Way, and is now accounted for in our new THEMIS dust model (6.3 below).

The main methodological limitations of these previous models were that we could not reliably interpret observations with a poor signal-to-noise ratio, and that there were potential noise-induced correlations and biases. 
These limitations originated in the simplicity of the statistical approach, which was the well-known least-squares minimization.

Motivated by the pioneering work of Kelly et al. (2012), we transformed our empirical SED model (Galliano et al., 2011) into a hierarchical Bayesian code. 
The principle of this approach consists in modelling simultaneously each SED, and the overall distribution of physical parameters (called a prior). 
This is done by sampling the probability distribution of the ensemble of parameters using a Markov chain Monte Carlo (MCMC) method.
This way, by making very unrestrictive assumptions about the functional interdependency of each physical parameters within our sample (sample of galaxies, or pixels in an image), we can remove false correlations, some biases, and exploit the information contained in poor-signal-to-noise ratio observations. 

\begin{figure}[htbp]
  \begin{tabular}{cc}
    \includegraphics[width=0.48\textwidth]{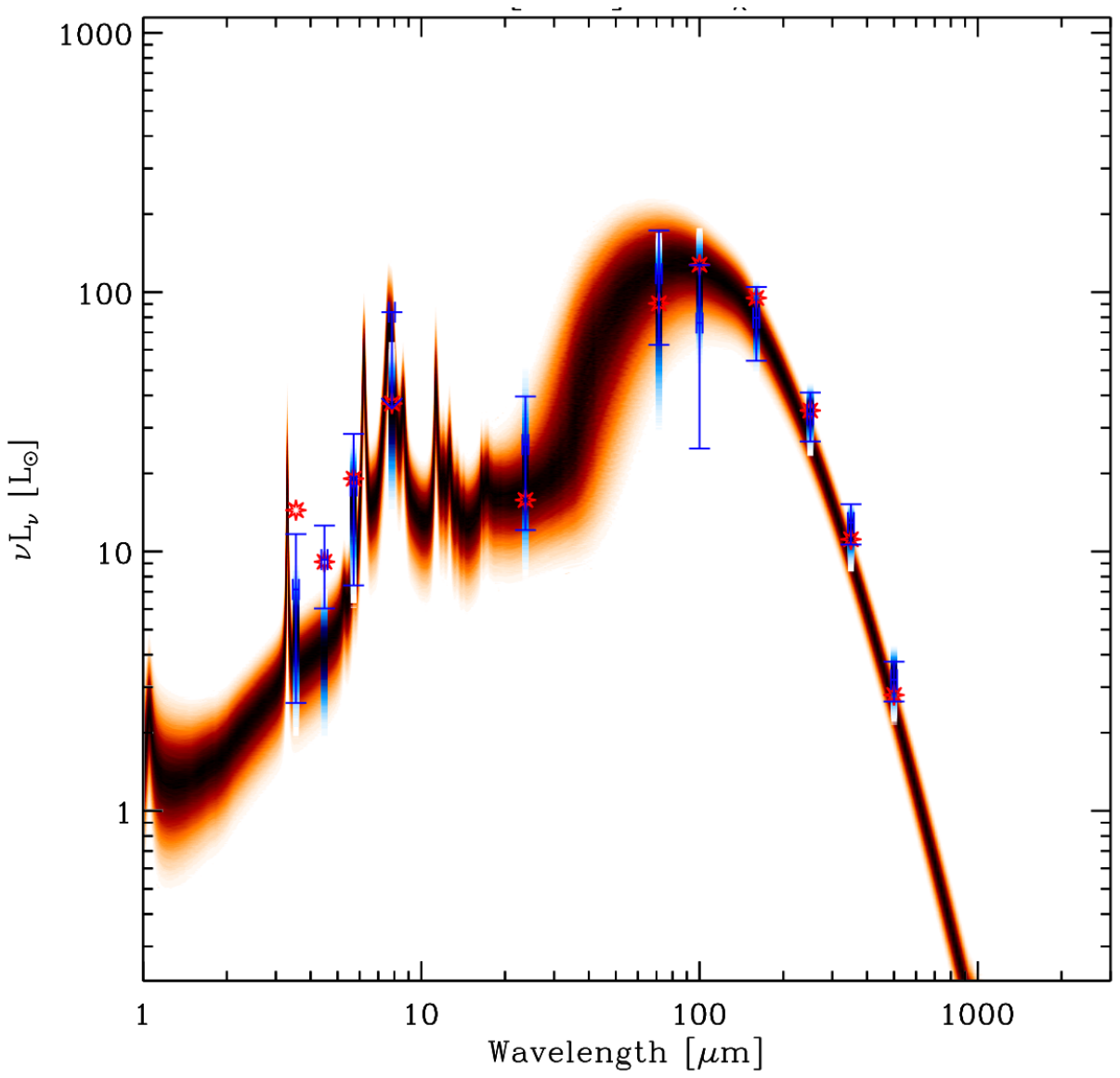} &
    \includegraphics[width=0.48\textwidth]{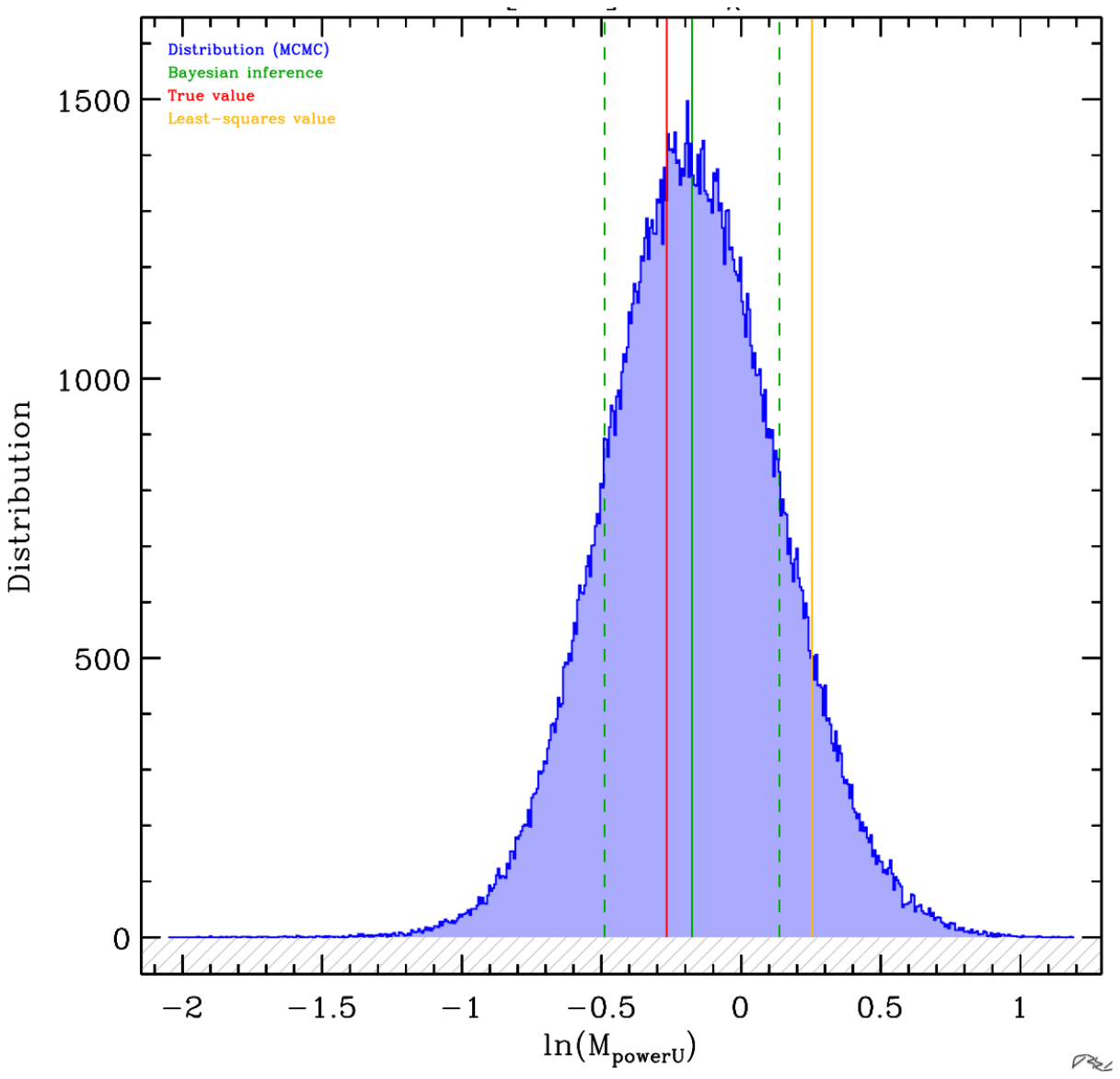} \\
  \end{tabular}
  \caption{{\sl Demonstration of the hierarchical Bayesian model} on simulated 
           data.
           These two panels show the results for one pixel of a simulation 
           containing 30.
           The left panel shows the simulated SED (red stars), with the 
           addition of noise and systematic uncertainties (blue error bars), at 
           a signal-to-noise level of $S/N=2.14$. 
           The red-to-black density shows the probability density of the SED, 
           inferred by HerBIE.
           The right panel shows the posterior probability distribution of the 
           derived dust mass (blue histogram), with its mean (green solid line) 
           and $\pm1\sigma$ (green dashed lines).
           The red solid line shows the true value, and the yellow solid line,
           the least-squares value.}
  \label{fig:HerBIE}
\end{figure}
Kelly et al. (2012) demonstrated this method, only for the simplest and analytical case of a modified black body. 
Our present hierarchical Bayesian dust SED model, called HerBIE (HiERarchical Bayesian Inference for dust Emission, Galliano, 2016) performs the complete inference of physical parameters, for an arbitrary large spectral cube with the linear combination of any of the following components:
\begin{itemize}
  \item an arbitrary number of modified black bodies;
  \item a uniformly-illuminated dust mixture;
  \item a dust mixture illuminated by a distribution of radiation fields;
  \item a stellar continumm;
  \item a free-free continuum;
  \item a synchrotron continuum.
\end{itemize}
The results are the full probability distribution of the physical parameters of each galaxy and/or each pixel in a 2-D image.
The model takes into account the correlated calibration uncertainties of each instrument.
An example is shown in Fig.~\ref{fig:HerBIE}.

Since this code is numerically intensive (large images may take several weeks to run), we have optimised it by pre-computing finely sampled templates of dust emission. 
We have developed a series of programs computing: the Mie theory for grain optical properties; stochastic heating of each species in a variety of heating conditions; integrated dust mixtures and synthetic photometry (including colour correction). 
We have built a database, using adaptive grids in wavelength, grain radius, grain temperature and radiation field intensity, in order to obtain precise results. 
The precision on the derived parameters is what will allow us to test the accuracy of each dust mixture. Full details of the HerBIE model and the tests that we have done to check it's robustness will be described in a subsequent paper (Galliano et al., in preparation).

\begin{figure}[htbp]
  \begin{tabular}{cc}
    \includegraphics[width=0.48\textwidth]{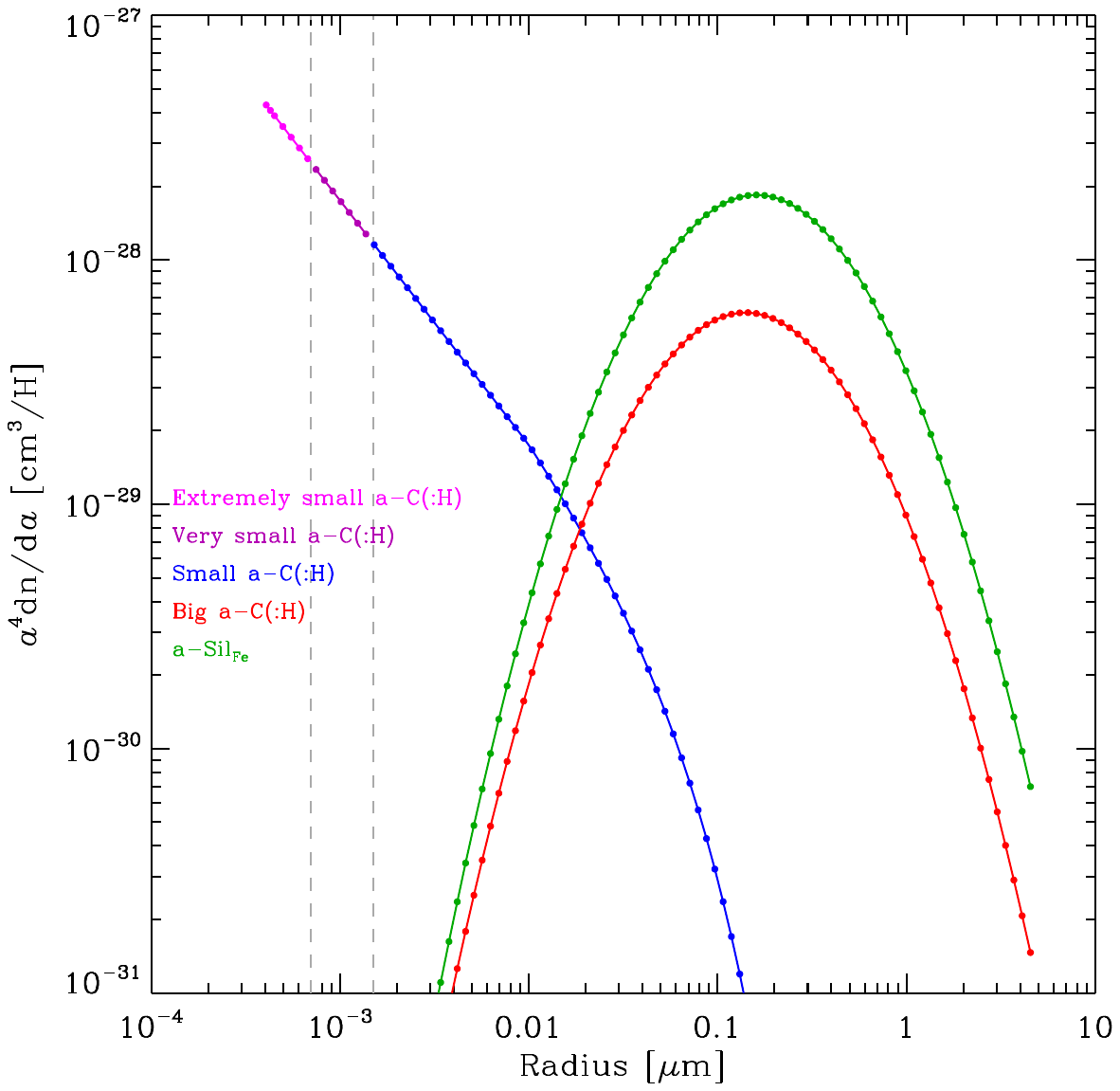} &
    \includegraphics[width=0.48\textwidth]{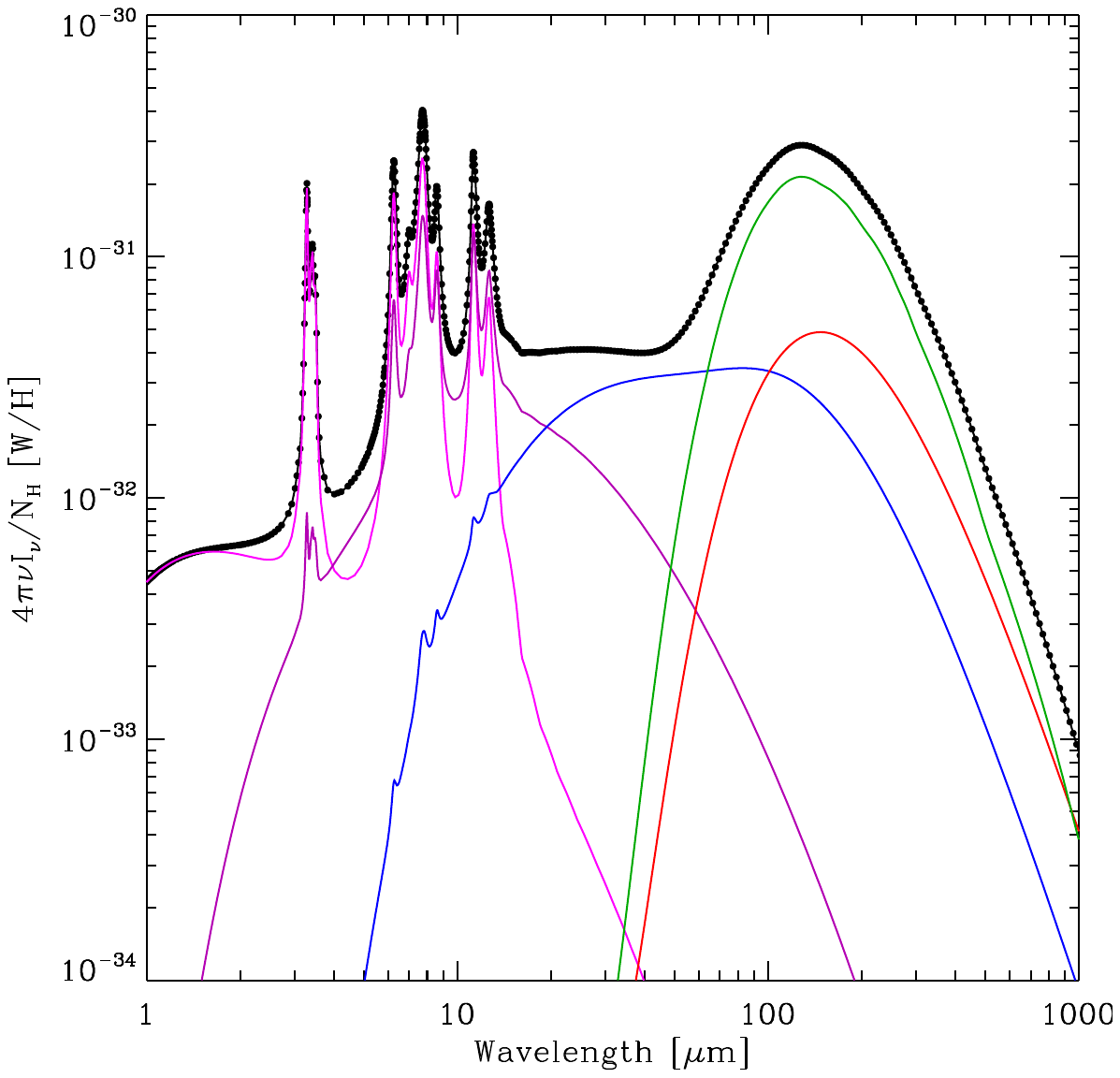} \\
  \end{tabular}
  \caption{{\sl Parameterization of THEMIS in HerBIE.}
           The left panel shows the size distribution of THEMIS, with the 
           our subdivision of the small a-C(:H) component.
           The right panel (same color code) shows the corresponding components 
           in emission.}
  \label{fig:sizedist}
\end{figure}
We have recently implemented the THEMIS dust model into HerBIE, using the newly obtained grain optical properties. 
The physics in the THEMIS code is different from that of the other dust mixtures we have previously been using: Zubko et al. (2004), Compiegneet al. (2011), Gallianoet al. (2011).
The main difference is in how small carbon grains are treated. 
Our previous dust mixtures assumed neutral and charged PAHs to reproduce aromatic features, and small graphite or amorphous carbons to account for the mid-IR continuum. 
The THEMIS dust model accounts for this ensemble of observables with hydrogenated amorphous carbons (HACs). 
We have found a way to parameterise the size distribution of these HACs, separating into several bins of sizes, in order to have a model with the same degree of freedom.
This is demonstrated in Fig.~\ref{fig:sizedist}. 
In subsequent papers we will compare the results of SED fitting using our previous dust mixtures and that of the THEMIS model.

\subsection{The dust extinction/emission connection, the sources of dust heating and the physical properties of galaxies - utilising radiative transfer models}

\begin{figure}
\centering
\vspace{-3.0cm}
\includegraphics[width=16.0cm]{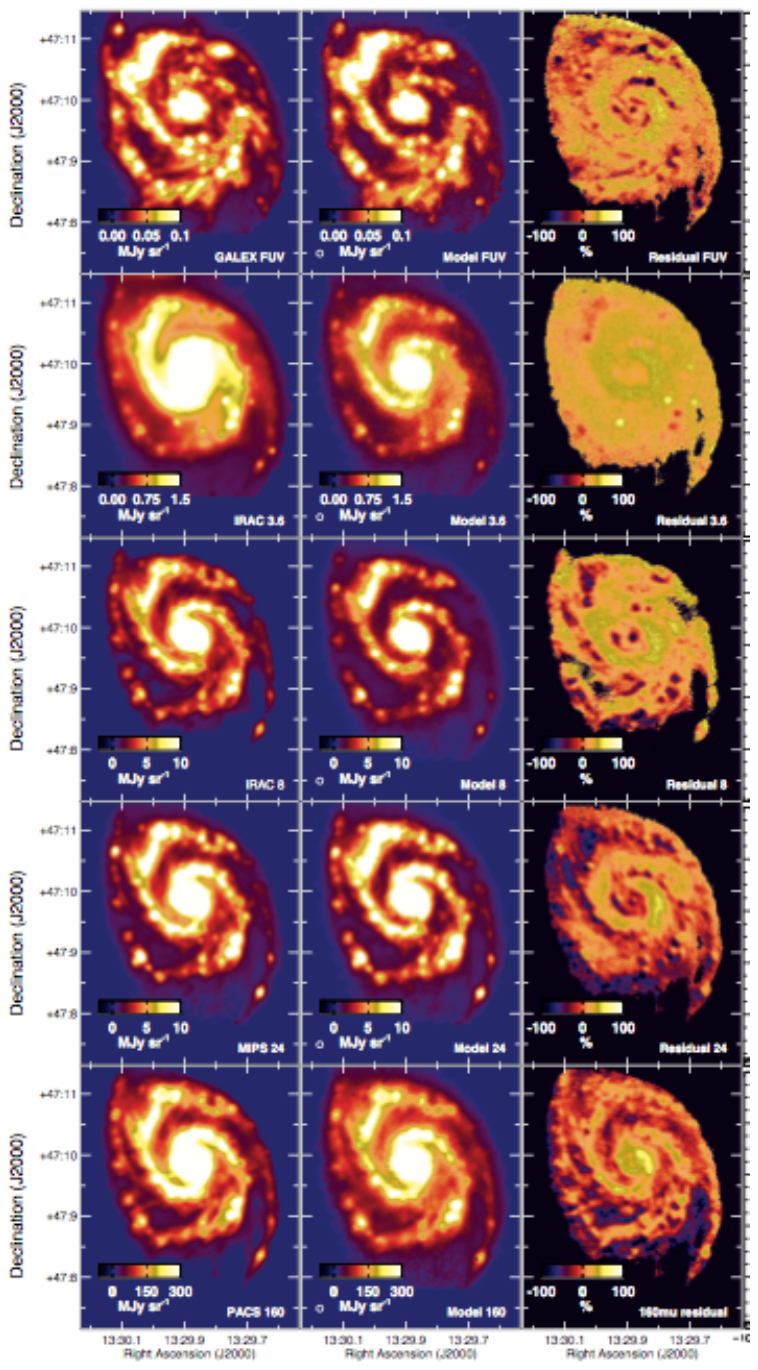}
\vspace{-2.0cm}
\caption{The observed (left column), modelled (2nd column), and residual (3rd column) images of M51 for the GALEX FUV (top row), IRAC 3.6 $\mu$m (2nd row), IRAC 8 $\mu$m (3rd row), MIPS 24 $\mu$m (4th row), and PACS 160 $\mu$m (5th row) wavebands. The observed images have been convolved to the resolution of the PACS 160 $\mu$m filter, i.e., the resolution of all input parameter maps in the radiative transfer code. The size of the PACS beam ($\approx$12 arcsec) is indicated in the bottom left corner of the model images. Image from De Looze et al. (2014). }
\label{fig:rad_trans}
\end{figure}

SED fitting (over the optical to sub-mm part of the spectrum), and in particular spatially resolved pixel-by-pixel SED fitting as described above, is an important step towards understanding the complex interplay between stars and dust in galaxies. However, such an analysis is by definition local, whereas non-local radiative transfer effects can play an important role. 

This is particularly relevant for studies of the dust heating mechanisms in galaxies. Current studies of dust heating are generally based upon correlations between properties of the dust (e.g.\ far-infrared flux or infrared colour ratios) and tracers of the interstellar radiation field (e.g.\ near-infrared emission) and star formation tracers (e.g. H${\alpha}$ or far-UV emission), and cannot take into account the non-local character of the heating (Boquien et al. 2011, Bendo et al., 2012, 2015, {\it Planck} Collaboration XXV 2015). However, many different sources can contribute to the heating of the dust in a galaxy, including the diffuse interstellar radiation field from evolved stars, young stars in star formation regions, and/or an active galactic nucleus. A prime example of nonlocal heating is the Sombrero Galaxy: De Looze et al. (2012a) have recently shown that the dust in this galaxy, which is located in a ring-like structure, is primarily heated by evolved stars in the massive bulge, rather than local sources in the ring itself. A similar scenario applies in the Andromeda galaxy (Groves et al. 2012, Viaene et al. 2016).

We therefore intend to apply full panchromatic radiative transfer modelling to an angular-diameter-limited subset ($D_{25}> 5$ arc min) of the DustPedia sample ($\sim 90$ galaxies). The goal of such a modelling approach is to quantify the relative importance of the dust heating sources, and simultaneously derive the 3D distribution and the spectral properties of the stellar populations and the interstellar dust in a galaxy. In this process, the effects of absorption, multiple scattering and thermal emission by dust are to be taken fully into account. In a radiative transfer modelling approach like ours, the dust is simultaneously studied in two complementary ways: through the extinction it causes in the UV and optical and through its direct thermal emission at FIR/sub-mm wavelengths. Assuming that photo-heating is the dominant mechanism to heat the dust, the absorbed and the thermally emitted energy budgets must be equal. 

We have previously developed a Monte Carlo radiative transfer code called SKIRT (Baes et al. 2003, 2011). While the code has been applied to a variety of objects including AGN tori (Stalevski et al. 2012), molecular clouds (Hendrix et al. 2015) and circumbinary discs (Deschamps et al. 2015), it has been specifically designed to model the interaction between starlight and interstellar dust grains within galaxies. It properly treats absorption and multiple anisotropic scattering by the dust, and offers a variety of simulated instruments for measuring the radiation field from any angle. It handles multiple dust mixtures and works with arbitrary 3D geometries. The code implements the common Monte Carlo optimisation techniques, such as peel-off at emission and scattering events, continuous absorption, and forced scattering, and includes a number of novel optimisation techniques (Baes 2008, Baes et al. 2011).

Up to now, SKIRT has mainly been used to model edge-on spiral galaxies (Baes et al. 2010, De Looze et al. 2012a, b, De Geyter et al. 2014, 2015, Mosenkov et al. 2016), as have several other radiative transfer codes (Xilouris et al. 1997, 1999, Bianchi 2007, MacLachlan et al. 2011, Schechtman-Rook et al. 2012). Edge-on spirals have the advantage that the extinction features and emission from dust can be easily distinguished along the line of sight. 

When self-consistent radiative transfer models are applied to edge-on galaxies, an inconsistency is usually found, in that the optical extinction generally underestimates the observed far infrared/sub-mm emission by a factor of about 3 (Bianchi et al., 2000, Popescu et al. 2000, 2011, De Geyter et al. 2015, Mosenkov et al. 2016). Two broadly different scenarios have been proposed to explain this difference: either there has been a significant underestimate of the dust emissivity (Alton et al. 2004, Dasyra et al. 2005), or the dust is significantly clumped so that it has a small effect on the extinction of the bulk of the starlight (Bianchi 2008, Baes et al. 2010, Saftly et al. 2015). Until recently, discriminating between these two scenarios has been a problem, due to the limited number of edge-on galaxies that had the required panchromatic data set available (particularly observations beyond 100$\mu$m that measure the bulk of the dust emission), the limitations of the available dust models, and the limited power of the radiative transfer models. We are confident that these problems will be solved using the DustPedia data, our new dust model (THEMIS), and our powerful radiative transfer code.

It is our intention to extend this modelling effort from edge-on galaxies to galaxies at all inclinations. This implies complications for the modelling: for edge-on spiral galaxies, the 3D model underlying the radiative transfer simulations can be constructed with analytical components (line-of-sight projection washes away the details). This approach has its limitations for (nearly) face-on galaxies. Therefore, as part of the DustPedia project, the SKIRT code has been improved so that it is able to cope with the more complex geometries that are required to model real galaxies. 
\begin{enumerate}
\item We have redesigned the general lay-out and structure of the SKIRT code. Inspired by standard software design principles and patterns, the latest version of SKIRT has a modular implementation that can be easily maintained and expanded (Camps \& Baes 2015). For example, different dust models can easily be added, as we have done with the THEMIS dust model (Jones et al. 2013). The latest version of SKIRT is publicly available through GitHub\footnote{\url{https://github.com/SKIRT/SKIRT}} and on the SKIRT website\footnote{\url{http://www.skirt.ugent.be/}}.
\item We have implemented an extended suite of input models that can be used for the distribution of stars and dust. This suite of models contains analytical components, numerically defined models, and 3D geometries based on the deprojection of observed images. An extensive set of 'decorators' is provided that can combine and alter these building blocks to more complex structures. This approach results in more code transparency and maintainability, and ensures that very complex models can easily be constructed out of simple building blocks (Baes \& Camps 2015). 
\item In order to allow the construction of complex models with a large dynamic range, a suite of advanced spatial grids have been incorporated, and different algorithms to trace photon packages through these grids have been compared (Saftly et al. 2013, 2014, Camps et al. 2013). 
\item The computational efficiency of SKIRT has been increased by introducing a new technique to accelerate the transfer of photons through the dust (Baes et al. 2016) and we have successfully introduced new parallelization strategies (Verstocken et al., in prep.).
\item FitSKIRT (De Geyter et al. 2013), a code specifically designed to fit SKIRT radiative transfer models to a set of optical images by means of genetic algorithms, has been upgraded. It can now include an arbitrary number of components for the stars and the dust, and the PSF convolution procedures are upgraded. Details can be found in Viaene et al. (2015) and Mosenkov et al. (2016).
\item We have developed a suite of pre- and post-processing tools for the radiative transfer simulations (PTS, the Python Toolkit for SKIRT\footnote{\url{http://www.skirt.ugent.be/pts/}}). These tools increase the automation of the galaxy modelling process, which is an important step towards the application of SKIRT to a larger set of galaxies.
\end{enumerate}

We intend to apply the updated SKIRT code to model a substantial number of large DustPedia galaxies, using a two-level approach. In a first stage, an initial model for the 3D distribution of stars and dust in a galaxy is determined, based on images at UV, optical and infrared wavelengths. This initial radiative transfer model is then refined by varying some of the parameters until it matches the observations. This technique was first successfully tested on M51 (see Figure~{\ref{fig:rad_trans}}) and subsequently applied to M31 (Viaene et al. 2016). 

The SKIRT radiative transfer model produces high-resolution simulated images of galaxies across the entire UV-submm wavelength range so that we can directly compare the models to the full suite of DustPedia imaging data. Moreover, the availability of a 3D model allows for the quantification of the contribution of the different stellar populations to the heating of the dust at every position in the galaxy, including non-local heating effects, for details, see De Looze et al. (2014) and Viaene et al. (2016).

The relevant physical parameters (in table format) and the complete simulated data cube (in FITS format) for each modelled galaxy will be provided on the DustPedia archive. This will provide access to anyone wishing to analyse the output of the radiative transfer simulation, and use it for his/her own purposes. Moreover, we will also make available all necessary data to re-run (and adapt, if desired) the SKIRT radiative transfer simulations. These data will include the maps of the distribution of stellar and dust components, the SKIRT parameter file (the so-called ski file, see Camps and Baes 2015), and the SKIRT version number used for the modelling. An explanation on how to use these data to simulate and explore the 3D radiative transfer model will also be provided.

\subsection{A new global physical model for cosmic dust (THEMIS)}

\begin{figure}
\centering
\includegraphics[width=17.0cm]{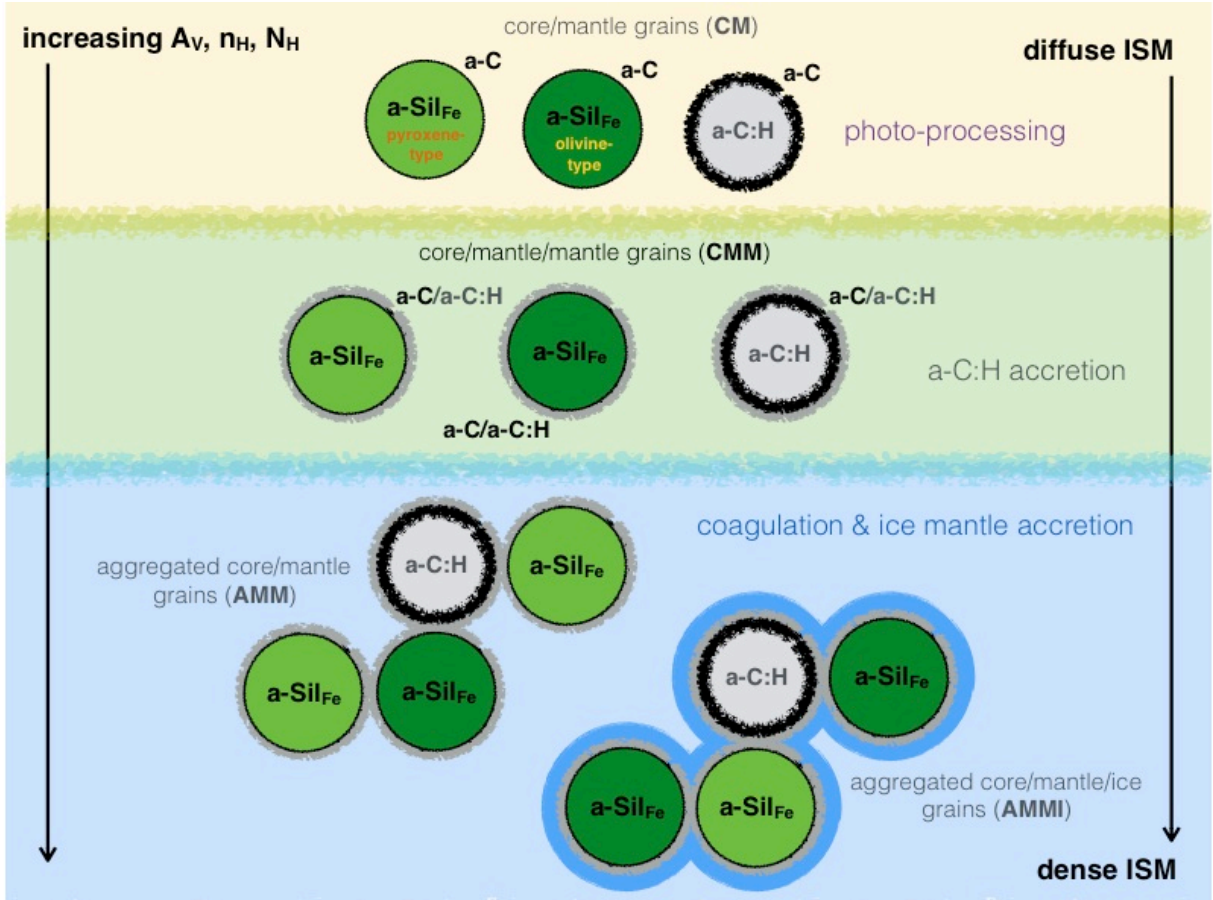}
\caption{Schematic view of the dust composition and stratification between diffuse ISM and dense molecular clouds in the THEMIS dust model. The major evolutionary processes acting on the dust in each region are indicated on the right, i.e., photo-processing, accretion and coagulation with increasing density and extinction from top to bottom (Jones et al. in prep 2016).}
\label{fig:themis}
\end{figure}

The follow-up to our work on the radiative transfer modelling of galaxies and the detailed analysis of their spectral energy distributions (SEDs) will be the interpretation of these data through our modelling of the physical processes operating on dust in the interstellar medium (ISM). We are currently developing and extending a self-consistent, unifying and global dust modelling framework called THEMIS (The Heterogeneous dust Evolution Model at the IaS). This framework is built upon the properties of amorphous hydrocarbon and amorphous silicate materials that have been measured in the laboratory (Jones et al. 2013, 2014, Kohler 2014).  THEMIS has been developed to include the evolution of these materials and the dust size distribution in the transition to dense regions (Kohler 2015, Ysard et al. 2016, Jones et al. 2016), in energetic regions (Bocchio et al. 2012, 2013, 2014), to explain H$_{2}$ formation in PDRs (Jones \& Habart 2015) and provide a link with elemental depletions and the diffuse interstellar bands (Jones 2013, 2014). Within the remit of the DustPedia project the modelling framework is being extended to the study of dust evolution in HII regions and PDRs (Figure~{\ref{fig:themis}}).

Dust within the THEMIS model is principally comprised of large carbon-coated amorphous silicate grains and small hydro-carbonaceous grains. The model was initially developed to investigate the nature of dust in the diffuse ISM and successfully explains the observed NIR-FUV extinction, IR-mm dust thermal emission and the shape of the IR emission bands. The model has also successfully confronted the latest available measures of the diffuse ISM dust extinction and emission in the Milky Way (Ysard et al. 2015; Fanciullo et al. 2015). The model correctly predicts the observed relationship between the observed E(B-V) extinction and the inferred mm opacity derived from the {\it Planck-HFI} observations (Ysard et al. 2015, Fanciullo et al. 2015).

The DustPedia global dust model will consider three key and distinct stages in the lifecycle of dust in interstellar media:
\begin{enumerate}
\item Dust sources - we will use the DustPedia data to constrain the relative contributions of evolved stars and supernovae to the silicate and carbonaceous dust budget. Evolved stars produce insufficient dust to explain current abundance and so there is increased interest in grain growth within the ISM. As part of DustPedia we are specifically considering the formation of dust in the dense, molecular ISM by coagulation and the accretion of atomic and molecular species from the gas phase.  A limitation of previous approaches to this problem is that they only empirically treat grain growth, because the process is not yet well understood physically. One key area where our approach differs is in the treatment of dust formation and growth through the effects of accretion and coagulation in dense molecular clouds. Our recent work (Jones et al. 2016; Ysard et al. 2016) shows that the accretion of carbon from the gas phase is extremely important locally. This work has given us some insight into how to better constrain dust formation in the ISM and we plan to build upon this in order to quantify it on a global galactic scale.
\item Dust evolution - we are further developing the dust properties (optical properties, composition, structure, etc.) in our existing models (Jones 2012a,b,c ; Jones 2013; Jones et al. 2014; Jones et al. 2013; Kohler et al. 2014; Jones et al. 2016; Ysard et al. 2016) with the aim of calibrating them on the dust SEDs in a variety of Milky Way cloud environments. These calibrated models can then be applied to the DustPedia data. Our aim here is to construct a unified dust model that is consistent with the various galactic types (elliptical, spiral, starburst, dwarf etc.) and with the expected spectral and spatial evolution of the dust populations within these galaxies. This will then feed back into our self-consistent radiative transfer models and SED fitting routines to predict the extinction, the observed far infrared/sub-mm emission and ultimately the dust mass in galaxies. These results can then be compared with those derived for other models (Zubko et al. 2004; Draine \& Li 2007;  Compiegne et al. 2011).
\item Dust sinks - dust is removed from the ISM principally through the effects of supernova-generated shock waves that erode it into its atomic constituents. However, it is now becoming clear that intense UV radiation fields, such as in photo-dissociation regions, can significantly affect the carbonaceous dust population. Thus, our models now include dust processing arising from sputtering, grain-grain collisions, electron, ion and UV irradiation in intense radiation fields, supernova shocks and in a hot coronal gas.  We therefore aim to quantify the dust destruction rate for given galaxy types and given galactic environments. 
\end{enumerate}

\subsection{Variations in dust emissivity/cross-section and the dust/gas/metallicity connection}

The DustPedia database, with its coverage of the peak of dust thermal radiation, offers a unique 
opportunity to study the dust emission properties in galaxies. We will use the DustPedia SEDs in the far infrared/sub-mm to derive the
dust emissivity (flux density per unit gas column density) in objects with available atomic
and molecular gas surface densities. Most of our galaxies have either 
global or resolved 21cm observations of atomic hydrogen, while molecular gas (CO) observations
are currently available for about 10\% of the sample. We will search for correlations 
within the data to help us predict molecular gas masses and so in combination with the atomic 
gas derive the total column density of hydrogen atoms. A major issue in the derivation of this
quantity is the poorly-known X$_{CO}$ conversion factor between CO observation and molecular gas 
column densities (for a review, see Bolatto, Wolfire \& Leroy, 2013).
However, for resolved galaxies with available {\it Herschel} data it has been shown that it is 
possible to derive $X_{CO}$ independently of the dust mass estimate and dust-to-gas mass ratio
(Sandstrom et a. 2013). We will pursue a similar method on our sample and study
correlations of  X$_{CO}$ with gas metallicity measurements. These are already available
as galaxy-integrated values for 260 objects of our sample (Hughes et al. 2013) and 
metallicity gradients are currently being derived for the largest targets (Hughes et al. 2015). Following the methodology of James et al. 
(2002, see also Clark et al. 2016)), the hydrogen mass content, the metallicity, and typical metal depletion patterns will 
be used to infer the mass of metals in dust and the dust-to-gas mass ratio. This will allow
us to express the emissivity in units of flux density per dust mass; and it will provide a 
comprehensive measure of the total amount of metals in a large sample of galaxies, a key 
parameter in galactic chemical evolution models. 

The emissivities derived using DustPedia galaxies will be studied as a function of morphology,
metallicity and (for larger objects) position within the galaxy. They will be compared with
the emissivity measurements in our Galaxy, and in particular with the FIR emissivity 
of high Galactic latitude dust which, together with the average extinction law in the optical, 
is one of the main constraints on dust grain models. Under the assumptions of 
dust heating by the local interstellar radiation field and optically thin emission
(well satisfied by the diffuse MW medium at high latitude), the modelled grain composition,
size distribution, and dust-to-gas mass ratio can be tuned to reproduce the MW emissivity.
Most current models are able to reproduce the far-infrared SED observed by the {\it FIRAS} instrument 
aboard the satellite COBE (Draine, 2003,  Compiegne et al. 2011, Jones et al. 2013), though updates are now necessary to explain the
recent measurements of a reduced emissivity obtained using {\it Planck} data ({\it Planck} collaboration XI, and results therein) and confirmed by {\it Herschel} data (Bianchi et al., in 
preparation).

From the emissivity, an average absorption cross section can be derived
under the simplifying assumptions of a simple power law dependence of
the cross section and a single mean temperature for all grains.
However simple, the model was found to provide results similar to those
of more complex physical models for dust emission (Dale et al 2012),
if most of the dust heating comes from an average interstellar
radiation field (Bianchi et al. 2013). However, a larger dynamical range
of heating conditions than predicted by this simple model could be present.
Also, the simultaneous derivation of the dust temperature and cross-section spectral index poses complex problems, given the fact that both contribute - in a degenerate way - to the shape of the modified blackbody. True changes in the spectral index were found in
M31 (Smith et al. 2012) and M33 (Tabatabaei et al. 2014), though the
results might be biased by the broadening of the SED caused by
the different temperatures attained by different grain materials
(Hunt et al. 2015).

With all these caveats in mind, we will attempt to derive the average dust cross-section by fitting with modified blackbodies the emissivity SEDs of the galaxies in our sample, mitigating the effects of the fitting degeneracy using the HerBIE code. Despite the limitations and possible pit-falls, this work will provide an independent estimate of the average cross-section, which will be compared with that resulting from our dust model (THEMIS). 

\subsection{The physical properties of dust in spiral galaxies and the dust distribution compared to the stars and the gas}

The determination of the content and spatial distribution of dust in spiral galaxies and the study of its effects on their observable properties has a rich and controversial history (Disney, Davies and Phillipps. 1989, Valentijn 1989, Davies and Burstein 1994, Xilouris et al. 1997, Bianchi et al. 2007, Xilouris et al. 1998, Driver et al. 2007, Baes et al. 2010, de Loose et al 2012, de Geyter et al. 2014, De Looze et al. 2014, Kirkpatrick et al. 2014). An improved knowledge is critical to both our understanding of the evolution of the cold interstellar medium and how to translate the observed properties of galaxies into physically meaningful quantities such as intrinsic luminosities, stellar spectral energy distributions and star formation rates. Despite the considerable recent progress in this field, the physical properties of the grains (structure, composition optical properties and spatial distribution) and how they may vary both within and between galaxies is still not fully understood, as witnessed by the energy budget problem described above. 

Our new radiative transfer models of spiral galaxies of all inclinations will enable us to investigate more fully the two competing solutions to the energy budget crisis - inhomogeneity and large scale structures in the star and dust distributions or changes in the dust emissivity. Preliminary results from our 3D clumpy Monte Carlo simulations indicate that a clumpy dust medium can help to restore the energy balance, but it is yet unclear whether it can generate the necessary dust mass to reconcile the independently derived optically and far infrared determined values (Bianchi 2008). The alternative scenario, advocated by Alton et al. (2004) and Dasyra et al. (2005), is that the energy balance in spiral galaxies can be restored if the dust emissivity at sub-mm wavelengths is about a factor four higher than that for diffuse Galactic dust. Indeed, high emissivities are found in models and observations of dust in Galactic dense clouds. However, it is not clear why cold dust (thought to be heated by the diffuse interstellar radiation field) should have properties typical of dense environments. Dust pervasive throughout the changing environments of spiral galaxies with varying physical properties has not previously been studied to any great depth and in a systematic way - something that the DustPedia data will now enable us to do. 
\begin{figure}
  \centering
  \includegraphics[width=14.5cm,angle=0]{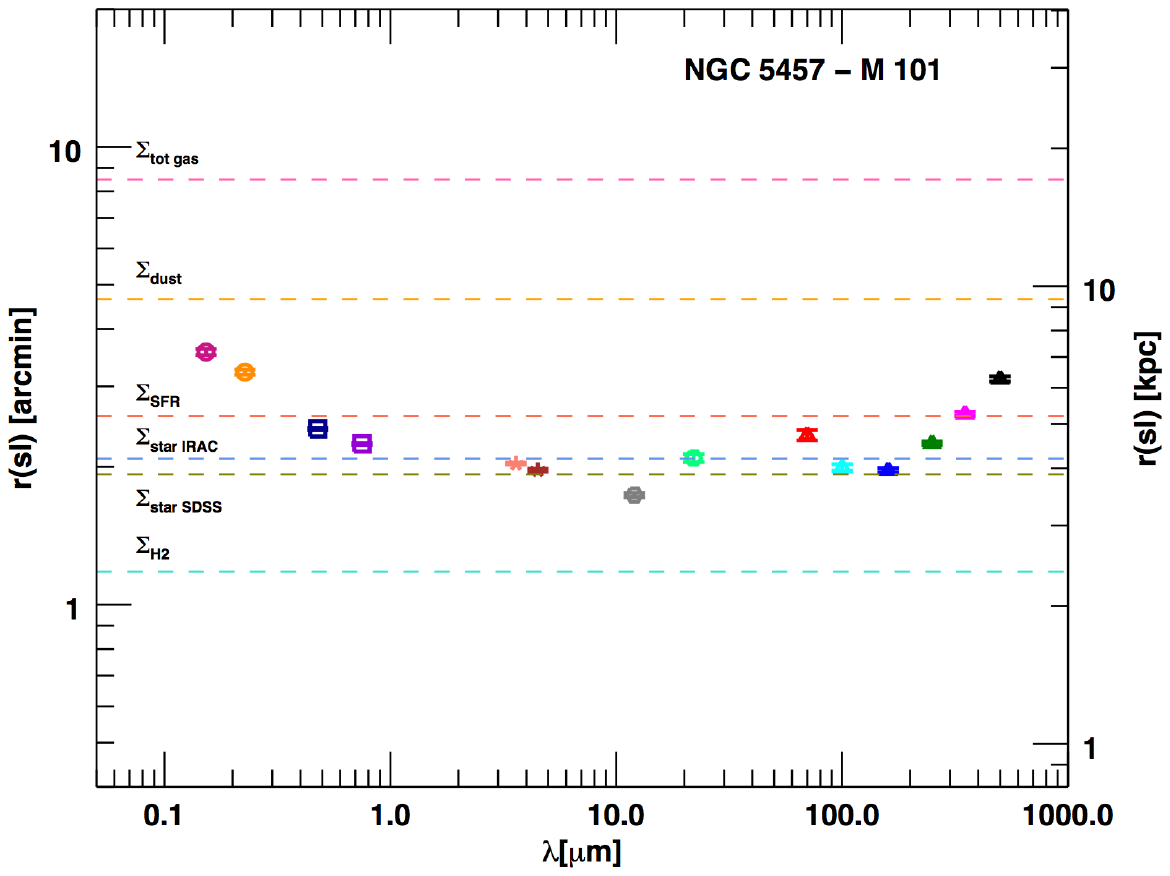}
  \vspace{-5.0cm}
  \caption{Exponential scalelengths of the surface brightness profiles of M101(NGC5457) in FUV and NUV {\it GALEX}, $g$ and $i$ SDSS, 3.6 and 4.5$\mu$m $\it{Spitzer}$, 
12 and 22$\mu$m {\it WISE}, 70, 160, 250, 350 and 500$\mu$m {\it Herschel}. The horizontal lines indicate the exponential scale length of the mass surface density of dust, molecular gas, total gas
and stars and that of the star formation rate density.}
\label{scalelengths}
\end{figure}
As an example of the sort of science projects we can undertake we have measured the radial exponential scale lengths of the stellar, gas and dust distributions of 18 nearby face-on spiral galaxies extracted from DustPedia sample (Casasola et al. 2016, in preparation). The connection between dust, gas, and stars is most often investigated through
spatially unresolved studies.
Although this is useful for characterising global disk properties, understanding intimately the structure of galaxies
requires resolved measurements.
Taking advantage of the unprecedented resolution of Herschel and the large angular size of many of the DustPedia sample galaxies, and using the available ancillary data from the UV to the sub-mm,
we can compute the azimuthally averaged profiles in each band, which we then fit to an exponential surface brightness distribution.
How the exponential scale lengths of each band vary and relate to each other for the galaxy M101(NGC5457) is show in Figure~{\ref{scalelengths}}. Also included for comparison is the total mass surface density distribution of the gas (HI+H$_{2}$), dust and stars. The clear result from this is just how much more quickly the surface density of stars falls with radius compared to that of the gas (mainly HI in the outer regions) and dust. If the source of the dust is existing galactic stars then there must be a mechanism to transport it to the outer regions of the disc.

\subsection{The origin and physical properties of dust in early type galaxies}

\begin{figure}
  \centering
  \vspace{-9.0cm}
  \includegraphics[width=16.5cm,angle=0]{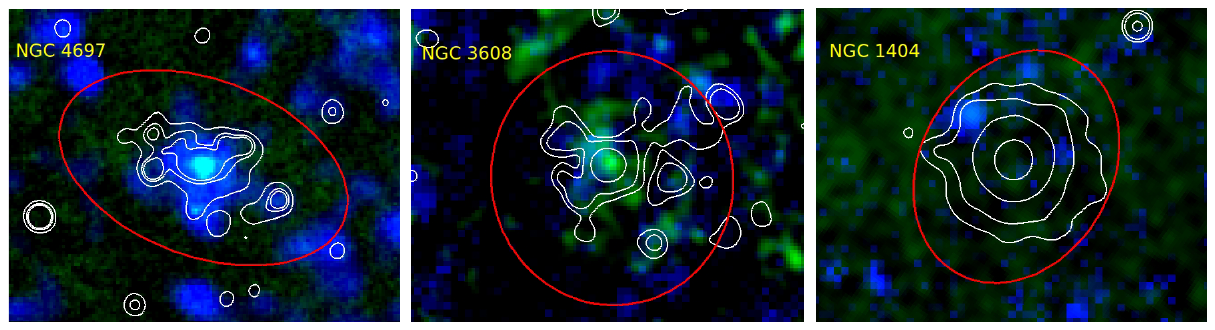}
  \vspace{-9.0cm}
  \caption{Three DustPedia early-type galaxies, NGC 4697, NGC 3608, and NGC 1404. Green is PACS 100 $\mu$m,
blue is SPIRE 250 $\mu$m, while the white contours indicate the X-ray emission of the hot gas as observed
by Chandra X-ray observatory. The red ellipse shows the B-band D$_{25}$ boundary of the
galaxy. It is evident that in many regions with excess of X-ray emission there is lack of dust.}
\label{fig9}
\end{figure}

Pinpointing the place of early type galaxies in galaxy evolution models remains an outstanding astronomical issue (Conselice 2008). One cosmological scenario has them forming early in the universe, leaving them as quiescent objects, evolving passively into old age. Other evidence supports the idea that they have formed from more recent galaxy-galaxy merger events and have some current star formation activity. Although early type galaxies were once thought to be devoid of gas and dust, previous far infrared and sub-mm observations have indicated dust masses some 10 - 100 times larger than those derived from the optical extinction alone (Knapp et al. 1989, Tsai and Mathews 1996). This implies that there must be a more diffuse component of cool dust present or again possibly that the far infrared emissivity maybe incorrect. Earlier mid-infrared observations with {\it ISOCAM} placed important constraints on the small end of the dust size distribution (Tsai and Mathews 1996). Many early types have excess mid-infrared emission that can be attributed to PAHs and very small grains. Recent {\it Herschel} observations have produced evidence for the presence of larger colder dust grains in some early types (see Figure~{\ref{fig9}}, Smith et al. 2012, di Serego Alghieri et al. 2013, Amblard et al. 2014, Agius et al. 2015, Lianou et al. 2016). However, to date conclusive experiments on the range of physical dust properties have been lacking because of the small samples with full spectral coverage from the UV to the far infrared/sub-mm. Determination of the physical properties and the quantity of dust in local early type galaxies, as well as its distribution relative to the gas and the stellar components, provide clues to its origin/fate and a link to the evolutionary history of the galaxy.  We will use the sub-sample of DustPedia early type galaxies (281 galaxies with Hubble Type index between -5 and 0) to measure the dust abundance, its emission spectrum and trace its spatial distribution. This analysis will allow us to characterise the physical properties of the dust residing in early-type galaxies and to investigate its evolution and survival in the harsh interstellar environment.

Additionally, the source of the dust in early type galaxies is a subject of much debate, with two major competing hypotheses (Smith et al. 2012):
\begin{enumerate}
\item Dust is produced by evolved stars - Over the course of their evolution, high redshift proto-elliptical galaxies process their gas and dust reservoirs into stars, polluting the Inter-Galactic Medium with enriched gas. After many Gyrs, the cold gas is exhausted, star formation all but ceases, the stellar populations fade away and the removal of dust from the interstellar medium through star formation ceases. The remaining dust is modified or destroyed via sputtering by the hot X-ray gas. Thus in early type galaxies, the creation of dust in supernovae and the destruction of grains in supernova-induced shocks is no longer taking place. As long as dust is not introduced from elsewhere, we essentially have a stellar system that currently produces dust in the atmospheres of its mass-losing evolved (AGB) stars and destroys it via sputtering in the hot gas. If the timescale for dust destruction is short (such that early type galaxies are now only left with dust produced by evolved stars), we only require the interpretation of the effects of sputtering to determine the elusive properties of dust produced in this way. If true the dust life-cycle is simpler and more easily understood than in an actively star forming galaxy. This is something we will be able to measure using the DustPedia database and then use these measurements to refine our models of dust input and output. If the dust has an internal origin, then its spatial distribution should also follow that of the stars, a scenario we can also test using this sample.
\item Dust is accreted during a galaxy merger - To try to identify those galaxies within our sample that may have undergone a recent merger, we will use images of optical obscuration as well as the dust emission because the merging process is expected to leave dust 'trails' across the galaxy. DustPedia data provides us with the appropriate resolution and wavelength coverage to look for the association of dust emission with these prominent extinction features. Sputtering timescales are typically $10^{6}$ to $10^{8}$ years while the dynamical timescale is greater than $10^{8}$ years - accreted dust should be sputtered before it is dispersed. Recent mergers should reveal themselves as strong extended dust features and not as the very small dust features often seen close to the centres of early type galaxies.
\end{enumerate}

Given the relatively uniform stellar population in early type galaxies one might expect that there would be small scatter in the dust-to-stars mass ratio if the dust was predominantly produced in evolved stars rather than deposited by the more stochastic merging process. Our studies will provide constraints to distinguish between these two scenarios because we can use the UV/optical emission to study the star and dust formation rates (input), the far infrared/sub-mm properties to measure the current dust mass (accumulation) and the x-ray emission to assess the strength of dust destruction via sputtering (output).
With their low level of ongoing star formation early type galaxies can provide a new and unique environment for the study of dust production, evolution and destruction. The properties of the dust can be compared with dust from other environments to give us a better understanding of processes in the interstellar medium of galaxies (see e.g. Agius et al. 2015). 

\begin{figure}
  \centering
  \vspace{0.0cm}  
   \includegraphics[width=16.5cm,angle=0]{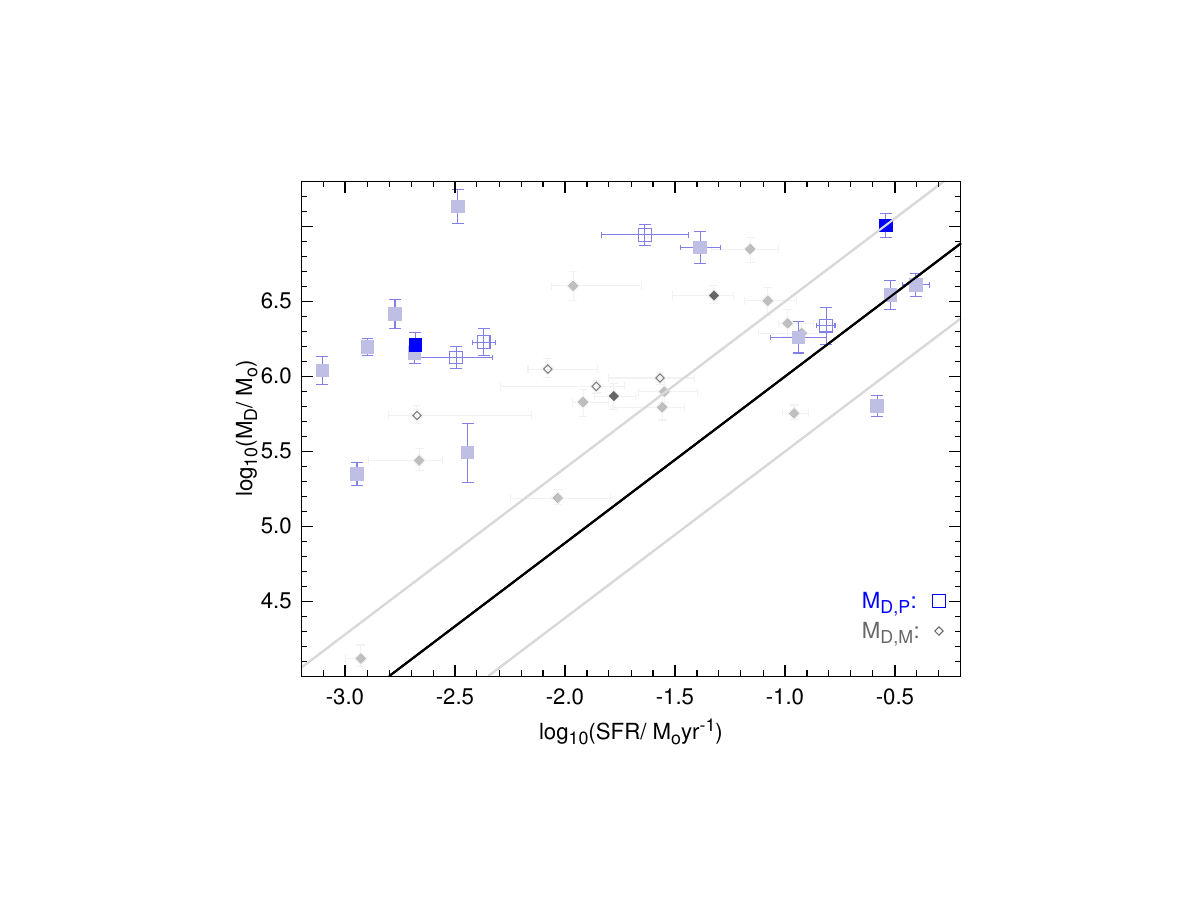}
  \caption{Dust mass as a function of SFR, using two different SED tools: MAGPHYS, shown in grey diamonds, and PCIGALE, shown in blue squares. Dark-coloured filled symbols indicate early type galaxies detected in both HI and CO, while those detected in only HI have light-coloured filled symbols; open symbols are ETGs detected in CO only. The relation found by da Cunha et al. (2010) is shown with the black solid line, with the intrinsic scatter indicated by the two thin grey-lines. Error bars represent the standard deviation of the probability distribution function. See Lianou et al. 2016 for more details.}
\label{fig10}
\end{figure}

We have already studied the dustier early type galaxies detected with Herschel and investigated how they differentiate from normal star forming galaxies. Figure~{\ref{fig10}} shows one such example where early type galaxies deviate from the relation defined by star-forming galaxies when placed in the SFR-dust mass plane (Lianou et al. 2016). The full DustPedia sample will allow us to investigate such scaling relations and compare early type galaxies with galaxies of other types. 

 \subsection{The nature of dust in star forming low metallicity (dwarf) galaxies}

Only in the local Universe can we detect far infrared emission from dwarf galaxies (Grossi et al. 2010, de Loose et al. 2010). 
Their importance is that they provide a unique opportunity to study star formation and processes in the interstellar medium 
in galaxies of very low metallicity, similar conditions to those that other galaxies would have experienced during their early history. 
One of the most interesting conundrums to recently come to light is the behaviour of their dust-to-gas mass ratios as a function 
of metallicity (Cannon et al. 2006, Galliano et al. 2005, Remy-Ruyer et al. 2013, 2014, 2015). On the one hand the high dust-to-gas 
ratios determined for some of these star forming dwarf galaxies (using the usual graphite dust to explain the emission properties) 
results in a dust-to-gas ratio too high considering the metallicities and the limit of the amount of metals possible to be incorporated 
into dust. On the other hand, for metallicity values lower than about 0.1 solar, the dust masses are too low compared to the observed gas 
mass. For galaxies with dust-to-gas ratios higher than reconcilable, the dust masses may be reduced by replacing the commonly used 
graphite with other more emissive grains, such as amorphous carbon - a solution currently suggested for the low metallicity dwarfs 
(Galliano et al. 2011). Hence, these galaxies offer the possibility of studying cosmic dust that may be quite different to that in our 
own Galaxy, thus opening up new avenues of investigation with regard to processes in the interstellar medium. From the DustPedia sample 
we can extract a sample ($\approx70$ galaxies) of star forming dwarfs to investigate this problem further. Fitting the spectral energy 
distribution provides us with the dust mass and data from the literature and our own correlations will provide us with the gas mass.

There is already some evidence from the spectral energy distribution that the constituents and physical make up of the dust is different for these galaxies when compared to others. Recently, Remy-Ruyer et al. 2015 confirmed the existence of warmer dust in a significant sample of low-metallicity
sources with their SED being broader and peaking at shorter wavelengths compared to more metal-rich system.
An important question is whether the dust emissivity is different? For example, there is evidence that some of 
these star forming dwarf galaxies show excess emission at 500$\mu$m (above that expected from a modified blackbody),
that is not seen in more massive galaxies (Grossi et al. 2010). As part of DustPedia we will combine the {\it Herschel} and {\it Planck} 
observations with those at shorter wavelengths from  {\it Spitzer} and {\it WISE} to investigate properties of the grains emitting at MIR 
wavelengths, where emission from PAHs and very small warm grains dominate Galliano et al. 2008).

There are also other important questions we can address with the DustPedia data:
\begin{enumerate}
\item How do metals become incorporated into dust?  
\item How efficient is this 
process? 
\item Are silicate dust grains and carbon dust incorporated into dust similarly? 
\item Is there differential depletion between the 
different dust components?  
\end{enumerate}
One of the most fundamentally important measurable quantities in galaxies that can be studied to this 
end is the behaviour of the dust-to-gas ratio as a function of a galaxy's metallicity.  DustPedia will contribute to this issue with the 
critical low metallicity component. Our progress in dust modelling and the detailed SED modelling in DustPedia will quantify 
accurately the reservoirs of the elements and the dust budget in low metallicity galaxies - a step only recently possible with the 
large mid-infrared to mm dust tracers now available.  
Such an approach is needed since simple modified blackbody models may lead to an underestimation of the dust mass by a 
factor two or three e.g. Remy-Ruyer et al. 2015.
Since gas masses in dwarf galaxies are vastly dominated by the atomic phase, 
the available HI masses will provide the bulk of the total gas reservoir for this study. Molecular hydrogen is difficult to measure 
in these low mass galaxies. In these dwarf galaxies, the deficit of CO line emission, which we take for granted as a viable tracer 
for molecular hydrogen, is reduced by the low metallicity and is also dissociated by the radiation field produced by these actively 
star forming galaxies. How much molecular gas remains undetected in the low metallicity (low Av) ISM has been difficult to quantify 
for many galaxies using the observations. To date more than half of the DustPedia low metallicity sample has already been or is 
currently being observed with ground-based telescopes (e.g. Schruba et al. 2012, Cormier et al. 2010 or with the {\it Spitzer} and {\it Herschel} 
spectrometers (Madden et al. 2012). Thus, studies using DustPedia with detailed modelling of the gas tracers, as is being attempted 
for a handful of low metallicity galaxies (e.g.,  Cormier et al. 2010, Lebouteiller et al. 2012) will quantify and 
characterise this elusive ``dark'' molecular component. 
 
\subsection{Cosmological implications}
\begin{figure}
\vspace{-6.0cm}
\centering
\includegraphics[width=17.0cm]{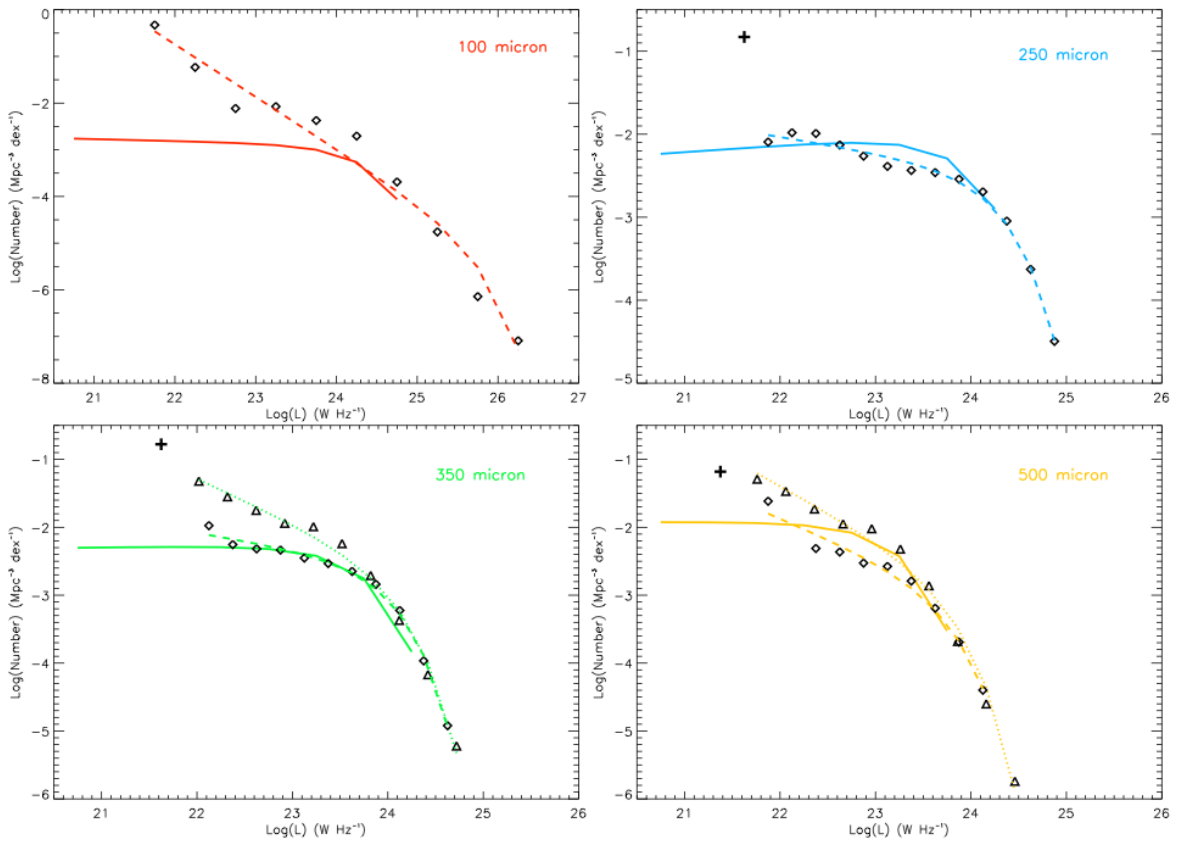}
\vspace{-7.3cm}
\caption{A comparison of luminosity functions at different wavelengths and over different environments (taken from Davies et al. 2014, Fig. 5)  . 100 micron - data points are from the {\it IRAS} survey of Sanders et al. (2003) and the dashed line is the Schechter function fit. The solid line is the Virgo cluster luminosity function normalised at $L^{*}$. 250 micron - data points come from Eales et al. (2015) and the dashed line is the Schechter function fit ignoring the lowest luminosity point marked by a cross. The solid line is the Virgo cluster luminosity function normalised at $L^{*}$. 350 micron - the diamond data points come from Eales et al. (2015) and the dashed line is the Schechter function fit ignoring the lowest luminosity point marked by a cross. The triangular data points come from Negrello et al. (2013) and the dotted line is the Schechter function fit. The solid line is the Virgo cluster luminosity function normalised at $L^{*}$. 500 micron - the diamond data points come from Eales et al. (2015) and the dashed line is the Schechter function fit ignoring the lowest luminosity point marked by a cross. The triangular data points come from Negrello et al. (2013) and the dotted line is the Schechter function fit. The solid line is the Virgo cluster luminosity function normalised at $L^{*}$.}
\label{fig:cosmology}
\end{figure}

To some extent it is probably true to say that we know more about the dust properties of distant galaxies than we do about those local to us. Galaxy surveys such as those undertaken using {\it IRAS} data (Saunders et al. 1990, Fairclough et al. 1986) and those in the sub-mm using, for example SCUBA (Eales et al. 2000) have identified populations of dust enshrouded galaxies producing most of their stellar radiation in the far infrared/sub-mm. These surveys have been further enhanced by the deep cosmological fields observed by {\it Herschel} (Oliver et al. 2012, Asboth et al. 2016). These predominately distant galaxies must have produced enormous amounts of cosmic dust over rather short periods of time and there is good evidence for an increasing co-moving dust mass density as we look back through cosmic history (Dunne et al. 2011). Today we observe locally the result of this evolution and to assess just how strong it has been we need to carry out a detailed analysis of the properties of local galaxies. Local galaxies as measured by DustPedia will provide a benchmark by which the more distant galaxy populations can be compared. For example one important measure of galaxy evolution is the change with time in the co-moving far infrared luminosity density. This is typically measured by integrating the far infrared luminosity function of galaxies at different redshifts. In Figure~{\ref{fig:cosmology}} we show and compare some recent determinations of far infrared luminosity functions (Davies et al. 2014). There are obviously differences between luminosity functions derived in different bands, using different instruments and for galaxies residing in different environments. We hope to address these issues with the DustPedia sample, by more precisely defining the very local (within 3000 km s$^{-1}$) luminosity functions and luminosity density. This far infrared luminosity density can be compared directly with that emitted by stars, taken from our SED and radiative transfer models. This will provide a "global" value for the fraction of radiation absorbed by dust and a "typical" optical depth for stellar photons. 

Using our derived luminosity functions and our dust model we can calculate the dust mass function and density of the local Universe and compare with the density of gas and stars (Davies et al. 2011, Panter et al. 2011) to derive the global baryon budget and to give dust-to-gas and dust-to-stars ratios for the local Universe.  Values for the dust mass density over much larger distance scales have previously been done in a statistical way (photometric redshifts) (Dunne et al. 2011), but never in the definitive way we are proposing here and not to the low galactic dust masses that we can achieve. The stellar, gas and dust (metals) mass fractions are linked via a chemical evolution model of galaxies (Davies et al. 2014 and references therein) and so we can make inferences about, for example, how much gas and/or metals have been lost by galaxies to the local inter-galactic medium or how they are currently being enriched by in falling gas. 

Another important cosmological measurement is that of the diffuse cosmic far infrared background (Puget et al. 1996) and how this relates to the galaxy population, for example can the source counts account for the background or is there some 'hidden' source of far infrared radiation (Viero et al. 2015). An important part of our analysis will be a measurement of the local luminosity density, which acts as a no evolution bench mark. This is something we will be able to do with the DustPedia data and particularly do this for the first time using the longer ($>250\mu$m) wavelengths provided by {\it Herschel}. 

Finally, Menard et al. 2010 have calculated the extent of dust around galaxies and the implication this has with regard to the dust optical depth of the Universe (see also Smith et al. 2016). Menard et al. did this by considering the extinction of quasars by galaxies. Using our radiative transfer models of resolved galaxies and extrapolating the derived dust density profiles we can repeat this measurement, but now using dust measured via its emission and our dust model, rather than by its direct optical extinction. 

\section{Conclusions}
The DustPedia project has primarily been funded to exploit the {\it Herschel} and {\it Planck} data archives that contain many observations of nearby galaxies. We have selected a {\it WISE} 3.4 $\mu$m sample of nearby galaxies and then looked for their counterparts in the {\it Herschel} and {\it Planck} Science Archives. From the archives we have been able to select 876 galaxies that have far infrared and/or sub-mm data. We have carried out our own data reduction of these galaxies and ingressed this data into the DustPedia galaxy database. We have then searched for auxiliary data for these galaxies in databases ranging from the ultra-violet to the near infrared. This has resulted in just over 25,000 images, with many of our sample galaxies observed in twenty or more bands, these have again been ingressed into the DustPedia database. We have carried out photometry on all the data so that calibrated images and global flux values are available for the whole multi-wavelength data set. 

We are developing modelling tools to interpret the observations. These include a new SED fitting tool (HerBIE) that uses a Bayesian approach, a Monte Carlo photon tracing radiative transfer model (SKIRT) of galaxies and a new physical model for the dust (THEMIS). 

We will use the data and the output from our models to explore the origins of cosmic dust, its evolution in the inter-stellar medium and its ultimate fate. On more global scales we will study the influence dust has on processes in the interstellar medium and how it affects our interpretation of the global properties of galaxies of various morphological types and galaxies within different environments.

The complete data sets will eventually be used to study the luminosity and mass functions of galaxies and how these can be related to issues regarding the evolution of galaxies through cosmic time.

\begin{center}
{\bf Acknowledgements}
\end{center}
J.Fritz acknowledges the financial support from UNAM-DGAPA-PAPIIT IA104015 grant, Mexico.

The {\it Herschel} spacecraft was designed, built, tested, and launched under a contract to ESA managed by the {\it Herschel/Planck} Project team by an industrial consortium under the overall responsibility of the prime contractor Thales Alenia Space (Cannes), and including Astrium (Friedrichshafen) responsible for the payload module and for system testing at spacecraft level, Thales Alenia Space (Turin) responsible for the service module, and Astrium (Toulouse) responsible for the telescope, with in excess of a hundred subcontractors.

PACS has been developed by a consortium of institutes led by MPE (Germany) and including UVIE (Austria); KU Leuven, CSL, IMEC (Belgium); CEA, LAM (France); MPIA (Germany); INAF-IFSI/OAA/OAP/OAT, LENS, SISSA (Italy); IAC (Spain). This development has been supported by the funding agencies BMVIT (Austria), ESA-PRODEX (Belgium), CEA/CNES (France), DLR (Germany), ASI/INAF (Italy), and CICYT/MCYT (Spain).

SPIRE has been developed by a consortium of institutes led by Cardiff University (UK) and including Univ. Lethbridge (Canada); NAOC (China); CEA, LAM (France); IFSI, Univ. Padua (Italy); IAC (Spain); Stockholm Observatory (Sweden); Imperial College London, RAL, UCL-MSSL, UKATC, Univ. Sussex (UK); and Caltech, JPL, NHSC, Univ. Colorado (USA). This development has been supported by national funding agencies: CSA (Canada); NAOC (China); CEA, CNES, CNRS (France); ASI (Italy); MCINN (Spain); SNSB (Sweden); STFC (UK); and NASA (USA).

This research has made use of the NASA/IPAC Extragalactic Database (NED) which is operated by the Jet Propulsion Laboratory, California Institute of Technology, under contract with the National Aeronautics and Space Administration. 

Funding for the SDSS and SDSS-II has been provided by the Alfred P. Sloan Foundation, the Participating Institutions, the National Science Foundation, the U.S. Department of Energy, the National Aeronautics and Space Administration, the Japanese Monbukagakusho, the Max Planck Society, and the Higher Education Funding Council for England. The SDSS Web Site is http://www.sdss.org/.
The SDSS is managed by the Astrophysical Research Consortium for the Participating Institutions. The Participating Institutions are the American Museum of Natural History, Astrophysical Institute Potsdam, University of Basel, University of Cambridge, Case Western Reserve University, University of Chicago, Drexel University, Fermilab, the Institute for Advanced Study, the Japan Participation Group, Johns Hopkins University, the Joint Institute for Nuclear Astrophysics, the Kavli Institute for Particle Astrophysics and Cosmology, the Korean Scientist Group, the Chinese Academy of Sciences (LAMOST), Los Alamos National Laboratory, the Max-Planck-Institute for Astronomy (MPIA), the Max-Planck-Institute for Astrophysics (MPA), New Mexico State University, Ohio State University, University of Pittsburgh, University of Portsmouth, Princeton University, the United States Naval Observatory, and the University of Washington.

This publication makes use of data products from the Two Micron All Sky Survey, which is a joint project of the University of Massachusetts and the Infrared Processing and Analysis Center/California Institute of Technology, funded by the National Aeronautics and Space Administration and the National Science Foundation.

We acknowledge the usage of the HyperLeda database (http://leda.univ-lyon1.fr). 

This publication makes use of data products from the Wide-field Infrared Survey Explorer, which is a joint project of the University of California, Los Angeles, and the Jet Propulsion Laboratory/California Institute of Technology, funded by the National Aeronautics and Space Administration.

Based on observations made with the NASA Galaxy Evolution Explorer. GALEX is operated for NASA by the California Institute of Technology under NASA contract NAS5-98034. 
\newpage

\begin{center}
{\bf References}
\end{center}
Agius et al., 2015, MNRAS, 451, 3815 \\
Ahn et al., 2012, ApJS, 203, 21 \\
Alton et al., 1998, A\&A, 335, 807 \\
 Alton et al., 2004, A\&A, 425, 109 \\
 Amblard et al., 2014, ApJ, 783, 135 \\
 Aniano et al. 2012, ApJ, 756, 138 \\
 Asboth et al., 2016, arXiv160102665 \\
 Auld et al., 2013, MNRAS, 428, 1880 \\
 Baes and Dejonghe, 2001, MNRAS, 326, 722 \\
 Baes et al., 2003, MNRAS, 343, 1081 \\
 Baes et al., 2008, MNRAS, 343, 1081 \\
 Baes et al., 2010, A\&A, 518, L39 \\
 Baes et al., 2011, ApJS, 196, 22 \\
 Baes M. and Camps P., 2015, Astronomy and Computing, 12, 33 \\
 Baes et al., 2016, A\&A, 590, A55 \\
 Bell E., 2003, ApJ, 586, 794 \\
 Bendo et al., 2012, MNRAS, 419, 1833  \\
  Bendo et al., 2015, MNRAS, 448, 135 \\
  Bennett et al., 2003, ApJS, 148, 97 \\
 Berriman et al., 2016, AAS, 22734813 \\
     Bianchi et al., 2000, A\&A, 359, 65 \\
     Bianchi, 2007, A\&A, 471, 765 \\
 Bianchi, 2008, A\&A, 490, 461 \\
 Bianchi, 2013, A\&A, 552, 89 \\
 Bianchi et al., 2016, MNRAS, submitted \\
     Bocchio et al., 2012, A\&A, 545, 124 \\
     Bocchio, M., Jones, A.P., Verstraete, L., Xilouris, E.M., Micelotta, E.R., Bianchi, S. 2013, A\&A 556, A6 \\
Bocchio, M., Jones, A.P., Slavin, J.D. 2014, A\&A, 570, A32 \\
Bolatto, Wolfire \& Leroy, 2013, ARA\&A, 51, 207 \\
Boquien et al., 2011, AJ, 142, 111 \\
Calzetti et al., 2007, ApJ, 666, 870 \\
Camps P. and Baes M., 2015, Astronomy and Computing, 9, 20 \\
Camps P., Baes M. and Saftly W., 2013, A\&A, 560, A35 \\
    Cannon et al., 2006, ApJ, 652, 1170 \\
    Clark et al., 2016, MNRAS, in press \\
     Clark et al., 2015, MNRAS, 452, 397 \\
    Compiegne et al., 2011, A\&A, 525, 103 \\
    Conselice, 2008, "Pathways Through an Eclectic Universe", ASP Conference Series, 390, 403 \\
    Cormier et al., 2010, A\&A, 518, 57 \\
    da Cunha E., Charmandaris V., Díaz-Santos T., Armus, L., Marshall J. A. and Elbaz, D., 2010,  A\&A, 523A, 78 \\
    Dale et al., 2001, ApJ, 549, 215 \\
    Dale et al., 2012, ApJ, 745, 95 \\
    Davies J. and Burstein D., 1994, Proc of the NATO ARW "The Opacity of Spiral Discs", NATO ASI series, Vol 469 \\
  Dasyra et al., 2005, A\&A, 437, 447 \\
 Davies J. et al., 2010, A\&A, 518, 48 \\
 Davies J., et al., 2011, MNRAS, 415, 1883 \\
  Davies J. et al., 2011, MNRAS, 419, 3505  \\
 Davies J. et al., 2012, MNRAS, 419, 3505 \\
 Davies J. et al., 2013, MNRAS, 428, 834 \\
 Davies J. et al., 2014, MNRAS, 438, 1922 \\
 Davies L. et al., 2016, arXiv:1606.06299 \\
 De Serego Alighieri et al., 2013, A\&A, 552, 8  \\
 De Geyter et al., 2013,  A\&A, 550, 74 \\
  De Geyter et al., 2014,  MNRAS, 441, 869 \\
    De Geyter et al., 2015,  MNRAS, 451, 1728 \\
De Jong et al. and the KiDS and Astro-WISE consortiums, 2013, Experimental Astronomy, 35, 25. \\
    De Looze et al., 2010, A\&A, 518, 54 \\ 
  De Looze et al., 2012a, MNRAS, 419, 895 \\
    De Looze et al., 2012b, MNRAS, 427, 2797 \\
  De Looze et al., 2014, A\&A, 571, 69 \\
  Deschamps et al., 2015, A\&A, 577, A55 \\
  Desert et al., 2008, A\&A, 481, 411 \\
  Disney M., Davies J. and Phillipps S., 1989, MNRAS, 239, 939 \\
  Draine, 2003, ARA\&A, 41, 241 \\
  Draine \& Li, 2007, ApJ, 657, 810 \\
Driver et al., 2007, MNRAS, 379, 1022 \\
  Dunne et al., 2003, MNRAS, 341, 589 \\
   Dunne et al., 2011, MNRAS, 417, 1510 \\
  Edge et al., 2013, ESO Msngr, 154, 32. \\
 Eales et al., 2000, AJ, 120, 2244 \\
    Eales et al., 2012, ApJ, 761, 168 \\
  Eales et al., 2015, MNRAS, 452, 3489 \\
  Edmunds M. and Eales S., 1998, MNRAS, 299, 29 \\
  Egami et al., 2010, A\&A, 518, 12 \\
  Eskew et al., 2012, AJ, 143, 139 \\
    Fairclough, 1986, MNRAS, 219, 1p \\
    Fanciullo et al., 2015, A\&A, 580, 136 \\
    Galametz et al., 2011, A\&A, 532, 56 \\
    Galliano et al., 2003, A\&A, 407, 159 \\
    Galliano et al., 2005, A\&A, 434, 867 \\
 Galliano et al., 2011, A\&A, 536, 88 \\
  Galliano et al., 2016, in preparation \\
 Gomez et al., 2010, A\&A, 518, 45 \\
 Green et al., 2015, ApJ, 810, 25 \\
 Griffin et al., 2010, A\&A, 518, 3 \\
  Grossi et al., 2010, A\&A, 518, 52 \\
  Groves B. et al., 2015, MNRAS, 426, 892 \\
  Hendrix T., Keppens R. and camps P., 2015, A\&A, 575, A110 \\
  James A. et al., 2002, MNRAS, 335, 753 \\
     Jones, 2012a, A\&A, 542, 98 \\
     Hughes et al. 2013, A\&A, 550, 115 \\
     Hughes et al. 2015, Proceedings of the IAU Symposium "Galaxies in 3D across the Universe", Volume 309, 320 \\
     Hunt L. et al., 2015, A\&A, 576, 33 \\
 Jones 2012b, A\&A, 540, 1   \\
 Jones 2012c, A\&A, 545, 2 \\
 Jones, A. P. 2013, A\&A, 555, A39 \\
Jones, A.P., Fanciullo, L., K\"{o}hler, M., Verstraete, L., Guillet, V., Bocchio, M., Ysard, N. 2013, A\&A, 558, A62 \\
Jones, A.P., Ysard, N., K\"{o}hler, M., Fanciullo. L., Bocchio, M., Micelotta, E., Verstraete, L., Guillet, G. 2014, Faraday Discussion Meeting 168, 313 \\
Jones, A.P., Habart, E, A\&A, 581, A92 \\
Jones, A.P., K\"{o}hler, M., Ysard, N., Dartois, E., Godard, M., Gavilan, L. 2015, A\&A, 588, 43 \\
 Kelly et al., 2012, ApJ, 752, 55 \\
 Kessler M. et al., 1996, A\&A, 315, 27  \\
 Kirkpatrick et al., 2014, ApJ, 789, 130 \\
 K\"{o}hler, M., Stepnik, B., Jones, A.P., Guillet, V., Abergel, A., Ristorcelli, I., Bernard, J.-P. 2012, A\&A, 548, A61 \\
 Knapp et al., 1989, ApJS, 70, 329 \\
Lawrence et al., 2013, ViZier online data catalogue:UKIDSS-DR9, LAS, GCS and DXS surveys \\
Lebouteiller et al., 2012, A\&A, 546, 94 \\
Lianou S., Xilouris E., Madden S. C. and Barmby P., 2016, MNRAS, 461, 2856 \\
MacLachlan J. et al., 2011, ApJ, 741, 6 \\
Madden S. et al., 2012, IAUS, 284, 141 \\
Magrini et al., 2011, A\&A, 535, 13 \\
Makarov et al. 2014, A\&A, 570, A13 \\
  Menard et al., 2010, MNRAS, 405, 1025 \\
  Meidt et al., 2014, ApJ, 788, 144 \
   Misiriotis and Bianchi, 2002, A\&A, 384, 866 \\
   Mosenkov A. et al., 2016, A\&A, in press \\
   Munoz-Mateos et al., 2011, ApJ, 731, 10 \\
   Negrello et al., 2013, MNRAS, 429, 1309 \\
   Neugebauer G. et al., 1984, ApJ, 278, L1 \\
   Nguyen et al., 2010, A\&A, 518, 5 \\
   Oliver et al., 2012, MNRAS, 424, 1614 \\
   Panter B., Jimenez R., Heavens A. and Charlot S., (2007), MNRAS, 378,
1550 \\
 Peebles and Nusser, 2010, Nature, 465, 565 \\
 Pilbratt et al., 2010, A\&A, 518, 1 \\
 Planck collaboration XXV, 2015, A\&A, 582, A28  \\
 Poglitsch et al., 2010, A\&A, 518, 2 \\
 Popescu et al. 2000, A\&A, 362, 138 \\
  Popescu C. and Tuffs R., 2002, MNRAS, 335, 41 \\
 Popescu et al. 2011, A\&A, 527, A109 \\
 Puget et al., 1996, A\&A, 308, L5 \\
  Remy-Ruyer et al., 2012, IAUS, 284, 149 \\
 Remy-Ruyer et al., 2013, A\&A, 557, A95 \\
  Remy-Ruyer et al., 2014, A\&A, 563, 31 \\
   Remy-Ruyer et al., 2015, A\&A, 582, 121 \\
   Rice et al., 1988, ApJS, 68, 91 \\
 Roussel et al., 2013, PASP, 125, 125, 1126 \\
 Saftly W. et al., 2013, A\&A, 554, A10 \\
 Saftly W., Baes M. and Camps P., 2014, A\&A, 561, A77 \\
 Saftly W. et al., 2015, A\&A, 576, A31 \\
 Sanders D. and Mirabel I., 1996, ARA\&A, 34 749 \\
 Sanders et al., 2003, AJ, 126, 1607 \\
 Sandstrom et al., 2013, ApJ, 775, 5 \\
  Saunders et al., 1990, MNRAS, 242, 318 \\
  Schechtman-Rook et al., 2012, ApJ, 746, 70 \\
  Schruba et al., 2012, AJ, 143, 138 \\
 Shetty et al., 2009a, ApJ, 696, 223 \\
  Shetty et al., 2009b, ApJ, 696, 676 \\
  Skrutskie et al., 2006, AJ, 131, 1163 \\
 Smith, 2012a, PhD thesis University of Cardiff \\
   Smith et al., 2012b, ApJ, 756, 40 \\
   Smith et al., 2012c, ApJ, 748, 123 \\
   Smith et al., 2016, MNRAS, in press (arXiv160701020) \\
   Sodroski et al., 1997, ApJ, 480, 173 \\
   Solarz A., Takeuchi T. and Pollo A., 2016, arXiv:1607.08747 \\
   Stalevski et al., 2012, MNRAS, 420, 2756 \\
   Tabatabaei et al., 2014, A\&A, 561, 95 \\
   Tielens A., 2005, "The Physics and Chemistry of the Interstellar Medium", Cambridge University Press, Cambridge, UK.  \\
 Tsai and Mathews, 1996, ApJ, 468, 571 \\
 Valentijn E., 1990, Nature, 346, 153 \\
 Viaene S. et al., 2015, A\&A, 579, 103 \\
 Viaene S. et al., 2016, A\&A, submitted \\
 Viero et al., 2015, ApJ, 809, 22 \\
 Werner M. et al., 2004, ApJS, 154, 1 \\
Whittet, D., "Dust in the Galactic Environment (2nd Edition)", IOP Series in Astronomy and Astrophysics \\
 Wright et al., 2010, AJ, 140, 1868 \\
  Xilouris et al., A\&A, 1997, 325, 135 \\
 Xilouris et al., A\&A, 1998, 331, 894 \\
  Xilouris et al., A\&A, 1999, 344, 868 \\
  Ysard, N., K\"{o}hler, M., Jones, A., Miville-Desch�nes, M.-A., Abergel, A., Fanciullo, L. 2015, A\&A, 577, A110 \\
Ysard, N., K\"{o}hler, M., Jones, A.P., Dartois, E., Godard, M., Gavilan, L. 2015b, A\&A,  588, 44 \\
Zubko et al., 2004, ApJS, 152, 211 \\

\end{document}